\documentclass[%
 reprint,
superscriptaddress,
 amsmath,amssymb,
 aps,
 prl,
]{revtex4-2}

\usepackage{graphicx}
\usepackage{dcolumn}
\usepackage{bm}
\usepackage{hyperref}

\usepackage{xcolor}
\renewcommand*{\vec}[1]{\boldsymbol{#1}}

\begin{document}

\preprint{APS/Meng Wen}

\title{Spin polarized proton beam generation from gas-jet targets by intense laser pulses}

\author{Luling Jin}
\affiliation{%
 Department of Physics, Hubei University, Wuhan 430062, China
}%

\author{Meng Wen}
\email{wenmeng@hubu.edu.cn}
\affiliation{%
 Department of Physics, Hubei University, Wuhan 430062, China
}%

\author{Xiaomei Zhang}
\email{zhxm@siom.ac.cn}
\affiliation{State Key Laboratory of High Field Laser Physics, Shanghai Institute of Optics and Fine Mechanics, Chinese Academy of Sciences, Shanghai 201800, China}

\author{Anna H{\" u}tzen}
  \affiliation{Peter Gr{\" u}nberg Institut (PGI-6), Forschungszentrum J{\" u}lich, Wilhelm-Johnen-Str. 1, 52425 J{\" u}lich, Germany}
  \affiliation{Institut f{\" u}r Laser- und Plasmaphysik, Heinrich-Heine-Universit{\" a}t D{\" u}sseldorf, Universit{\" a}tsstr. 1, 40225 D{\" u}sseldorf, Germany}
  
  \author{Johannes Thomas}
  \affiliation{Institut f{\" u}r Theoretische Physik I, Heinrich-Heine-Universit{\" a}t D{\" u}sseldorf, Universit{\" a}tsstr. 1, 40225 D{\" u}sseldorf, Germany}

\author{Markus B{\" u}scher}
  \affiliation{Peter Gr{\" u}nberg Institut (PGI-6), Forschungszentrum J{\" u}lich, Wilhelm-Johnen-Str. 1, 52425 J{\" u}lich, Germany}
  \affiliation{Institut f{\" u}r Laser- und Plasmaphysik, Heinrich-Heine-Universit{\" a}t D{\" u}sseldorf, Universit{\" a}tsstr. 1, 40225 D{\" u}sseldorf, Germany}

 \author{Baifei Shen}
 \affiliation{State Key Laboratory of High Field Laser Physics, Shanghai Institute of Optics and Fine Mechanics, Chinese Academy of Sciences, Shanghai 201800, China}
 \affiliation{Department of Physics, Shanghai Normal University, Shanghai 200234, China}

\date{\today}

\begin{abstract}
A method of generating spin polarized proton beams from a gas jet by using a multi-petawatt laser is put forward. With currently available techniques of producing pre-polarized monatomic gases from photodissociated hydrogen halide molecules and  petawatt lasers, proton beams with energy \mbox{$\gtrsim 50$~MeV} and $\sim 80 \%$ polarization are proved to be obtained. Two-stage acceleration and spin dynamics of protons are investigated theoretically and by means of fully self-consistent three dimensional particle-in-cell simulations. Our results predict the dependence of the beam polarization on the intensity of the driving laser pulse. Generation of bright energetic polarized proton beams would open a domain of polarization studies with laser driven accelerators, and have potential application to enable effective detection in explorations of quantum chromodynamics.  
\end{abstract}

\maketitle


The polarization of a beam describes the collective spin state for an ensemble of particles. Since polarized particle beams play important roles in solving a wide variety of scientific and medical problems~\cite{Gentile:2017:RevModPhys.89.045004, Safronova:2018:RevModPhys.90.025008,MoortgatPick:2005cw}, the generation~\cite{Mane:RPP:2005} and  measurement~\cite{Adlarson:2014:PhysRevLett.112.202301} of polarization observables has a flourishing tradition at conventional particle accelerators.  In particular, spin polarized  proton beams enable key measurements in explorations of the quantum-chromodynamics landscape~\cite{Adamczyk:2016:PhysRevLett.116.132301, Aschenauer:RPP:2019}. High luminosity is required in such experiments for both high energy colliders to solve the outstanding proton spin puzzle~\cite{Florian:2014:PhysRevLett.113.012001,Alexandrou:2017:PhysRevLett.119.142002} and low energy colliders to extend Standard Model tests~\cite{Androic2018}. However, the intensity of the polarized beams is generally limited to several hundred milliamperes~\cite{Accardi2016}, which is very difficult to be further increased at a traditional accelerator~\cite{Mane:RPP:2005}. 
Although the beam quality of laser driven protons are not yet competing to the traditional accelerators, the beam intensity can be increased by several orders of magnitude.  Therefore, investigation of spin effects of protons in laser accelerators becomes timely and interesting, because it enables new low-cost, compact laser accelerators of polarized proton beams. 

With the advent of ultra-intense lasers up to \mbox{$\sim10$}~PW~\cite{Yanovsky:OE:2008,Guo:OE:2018,Yoon:OE:2019}, plasma accelerators are now capable of providing beams with energies of almost hundred MeV per unit charge~\cite{Wagner:2016:PhysRevLett.116.205002,Kim::POP:2016, Zhang:2017:PhysRevLett.119.164801, Hilz:NatC:2018,Higginson:NC:2018, Ma:2019:PhysRevLett.122.014803, Kraft:PPCF:2018}. There have been multiple theoretical works how to generate polarized beams in laser driven accelerators~\cite{huetzenHPLSE19, Buscher:IJMPA:2019,Li:2019:PhysRevLett.122.154801,Wen:2019:PhysRevLett.122.214801,Wu:2019:PhysRevE.100.043202,Thomas:2020}. These show that underdense plasmas provide an unique opportunity for generation of polarized beams already at existing laser facilities, thanks to recent experimental progress of producing polarized atomic hydrogen gases with densities above $10^{19}~\mathrm{cm^{-3}}$~\cite{Sofikitis:2018:PhysRevLett.121.083001}. Although such prepolarized gases can be used in principle to produce multi-MeV proton beams~\cite{Shen:2007:PhysRevE.76.055402, Zhang:NJP:2014,Wan:2019:PhysRevAccelBeams.22.021301, Engin:PPCF:2019}, an intuitive and optimized acceleration regime and a comprehensive understanding of the acceleration mechanisms is still pending. 

\begin{figure}[t]
    \includegraphics[width=8.5cm]{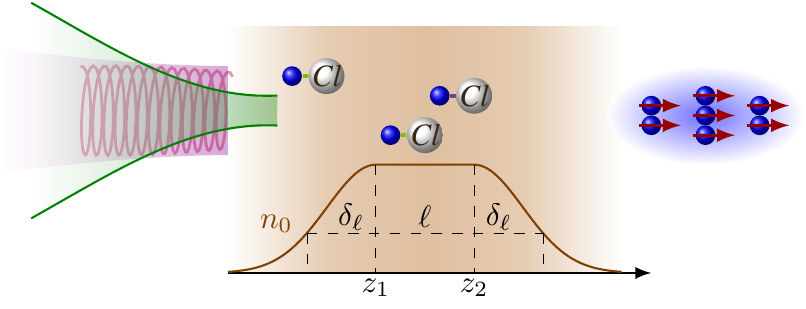}
    \caption{A schematic diagram showing laser acceleration of polarized protons from a dense hydrogen chloride gas target (brown).  Molecules are initially aligned along $z$-axis via a weak optical laser.  Blue and white balls represent nuclei of hydrogen and chlorine atoms, respectively. Before the acceleration driven by an intense laser beam (indicated by the green area), a weak circularly polarized UV laser (purple area) is used to generate the polarizes atoms along the longitudinal direction via molecular photodissociation. The brown curve indicates the initial density distribution of the gas-jet target. The polarized proton beam is shown on the right (blue) with arrows (red) presenting the polarization direction.}
    \label{fig:1}
\end{figure}

In this Letter, we report the first demonstration of laser driven polarized proton acceleration by means of full three-dimensional (3D) Particle-In-Cell (PIC) simulations using a basic scenario with available unltra-intense lasers and pre-polarized gases. The schematic diagram (see also in \cite{huetzenHPLSE19, Wen:2019:PhysRevLett.122.214801}) 
 is shown in Fig. \ref{fig:1}. The laser system comprises a weak circularly polarized UV laser to generate spin polarized atoms via photodissociation, and an intense laser to accelerate protons. The accelerated proton beam can be depolarized via asynchronous spin precession in inhomogeneous electromagnetic fields of laser driven plasmas. Using a gaseous HCl target with molecular density $\sim 10^{19}~\mathrm{cm^{-3}}$ and a 1.3 petawatt (PW) laser as an example, the energy of accelerated polarized protons can be as high as 50 MeV, with a beam polarization above 80 \%. The corresponding phase space distribution and the spin spread of accelerated protons are shown in Fig. \ref{fig:2}. As will be seen below, the energy increases with the power of the driving laser, while the polarization of the beam is almost preserved.

\begin{figure}[t]
    \includegraphics[width=8.5cm]{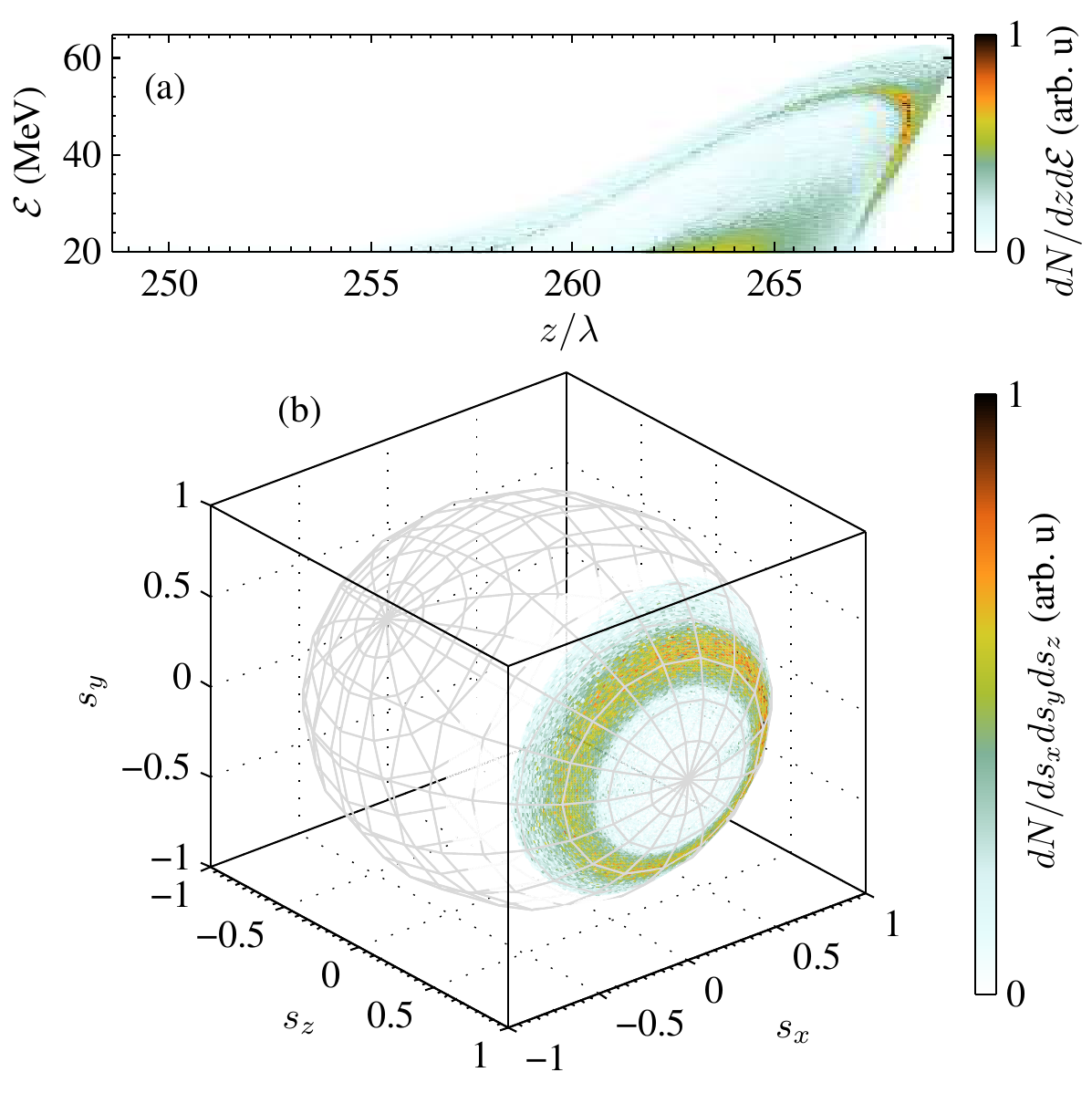}
    \caption{Snapshots at $t=330\lambda/c$ of: (a) phase-space distribution, and (b) spin spread of protons with energy $\mathcal{E}\geqslant 20$ MeV on the Bloch sphere. Simulation parameters can be found in the text.}
    \label{fig:2}
\end{figure}

We have proposed a related conceptual setup in \cite{huetzenHPLSE19}, however,  several criteria have to be satisfied for the success of laser induced spin-polarized beam acceleration. First, the parameters assumed in the simulations can be realized in experiments. Second, protons are accelerated effectively in the setup with a polarized/polarizable target. Third, a high degree of polarization of the accelerated protons is retained. The first criterion is met by adopting currently available PW lasers and a gas jet of polarized atoms with hydrogen density $\sim 10^{19}~\mathrm{cm^{-3}}$. The latter can be built via laser photodissociation with circularly polarized UV laser pulses (see Fig. \ref{fig:1})~\cite{Sofikitis:2018:PhysRevLett.121.083001}. After the molecules are dissociated to polarized atoms, the polarization is transferred from electrons to nuclei within a few 100 ps and oscillates between electrons and nuclei afterwards. Experiments should be designed such that the polarized gas is fully ionized by the PW laser when the hydrogen nuclei acquire a high degree of polarization. The delay between the UV laser and the driving pulse for ion acceleration should therefore be carefully controlled, which is accessible by splitting one laser beam~\cite{Wen:2019:PhysRevLett.122.214801} or by combining different laser systems~\cite{huetzenHPLSE19}. The remaining two criteria of efficient proton acceleration and small spin smearing under current available experiment conditions will be demonstrated below with 3D-PIC simulations.

Acceleration and spin precession of protons are investigated employing the EPOCH code ~\cite{Arber:epoch:2015hc} which has been extended by spin effects~\cite{Zamanian:2010:POP, Wen:2016:SP, Wen:2017:PhysRevA.95.042102}. The spin state of a charged particle with kinetic momentum $\vec p$ and velocity $\vec v$ is characterized classically by the unit vector $\vec{s} = (c_1^*c_2+c_1c_2^*,~
i c_1c_2^*-ic_1^*c_2,~\left|c_1\right|^2-\left|c_2\right|^2)$, with $c_1 \left|\uparrow\right> + c_2 \left|\downarrow\right>$ denoting the arbitrary spin state on the Bloch sphere of Fig. \ref{fig:2}(b) and $\left|\uparrow\right>$ a spin direction parallel to the $z$-axis. Furthermore, the Landau-Lifshitz equation $d\vec p/d t=\vec F_L + \vec F_R$ is applied to calculate the motion of electrons in the radiation-dominated regime driven by PW lasers~\cite{Landau:Lifshitz}. Here $\vec F_L=-q(\vec E + \vec v \times \vec B)$ is the Lorentz force and $\vec F_R \approx q^4/(6\pi \varepsilon_0m^2c^5)[(\vec v \cdot \vec E /c)^2-(\vec E + \vec v\times \vec B)^2] \gamma^2 \vec v$ is the Landau-Lifshitz force, with $\gamma$ the
relativistic Lorentz factor,  $c$ the speed of light in vacuum, $m$ and $q$ the electron mass and charge, respectively.

The 3D-PIC simulations are performed with a moving box of size $100\lambda\times 80\lambda\times 80\lambda$ represented by a $1000\times 400\times 400$ grid at speed of light $c$ and a total pseudoparticle number of $8\times10^8$. The laser pulse for proton acceleration ( green area in Fig. \ref{fig:1}) with a bi-Gaussian envelope $\vec a = \vec e_x a_0\exp[-(t-z/c+z_0/c)^2/\tau_0^2-r^2/w_0^2]$ and a focal position $z_0$ propagates along the $z$-axis, where the laser amplitude $a_0=25$, focal radius $w_0=10\lambda$, wavelength $\lambda = 800$ nm, and temporal duration $\tau_0=10\lambda/c$.  In order to reach a high acceleration efficiency, the laser pulse is focused to the left boundary of the gas target $z_0=z_1=50\lambda$. The target is assumed to be a fully ionized plasma of HCl gas, with all protons initially polarized along the $z$-axis $\eta_0=\left<\vec s_0\right>\cdot \vec e_z=1$. Its density profile, as shown by the brown curve in Fig. \ref{fig:1}, is a uniform plateau with steep edges~\cite{Semushin:RSI:2001, Schmid:RSI:2012, Lorenz:2019:MRE, Engin:PPCF:2019}. The length of the target is $\ell=z_2-z_1=200\lambda$. The density gradient at the edge is expressed as $n=n_0\exp[-(z-z_i)^2/\delta_\ell^2]$ with $\delta_\ell=5\lambda$, where $i=1$ and $2$ correspond to the left and right boundaries of the target, respectively. In the following simulations, gas jets with electron density $n_0=0.36n_c $ are applied where $n_c=\varepsilon_0 m\omega^2/e^2$ is the critical plasma density, $\varepsilon_0$ the vacuum permittivity, and $\omega=2\pi c/\lambda$ the laser frequency. One should note that the proton density $n_p$ is lower than the plasma density, i.e., $n_p=n_0/(Z_H+Z_{Cl})  \approx 3.48 \times 10^{19}~\mathrm{cm^{-3}}$, with charge numbers of hydrogen $Z_H=1$ and chlorine $Z_{Cl}=17$.

\begin{figure}[t]
    \includegraphics[width=8.5cm]{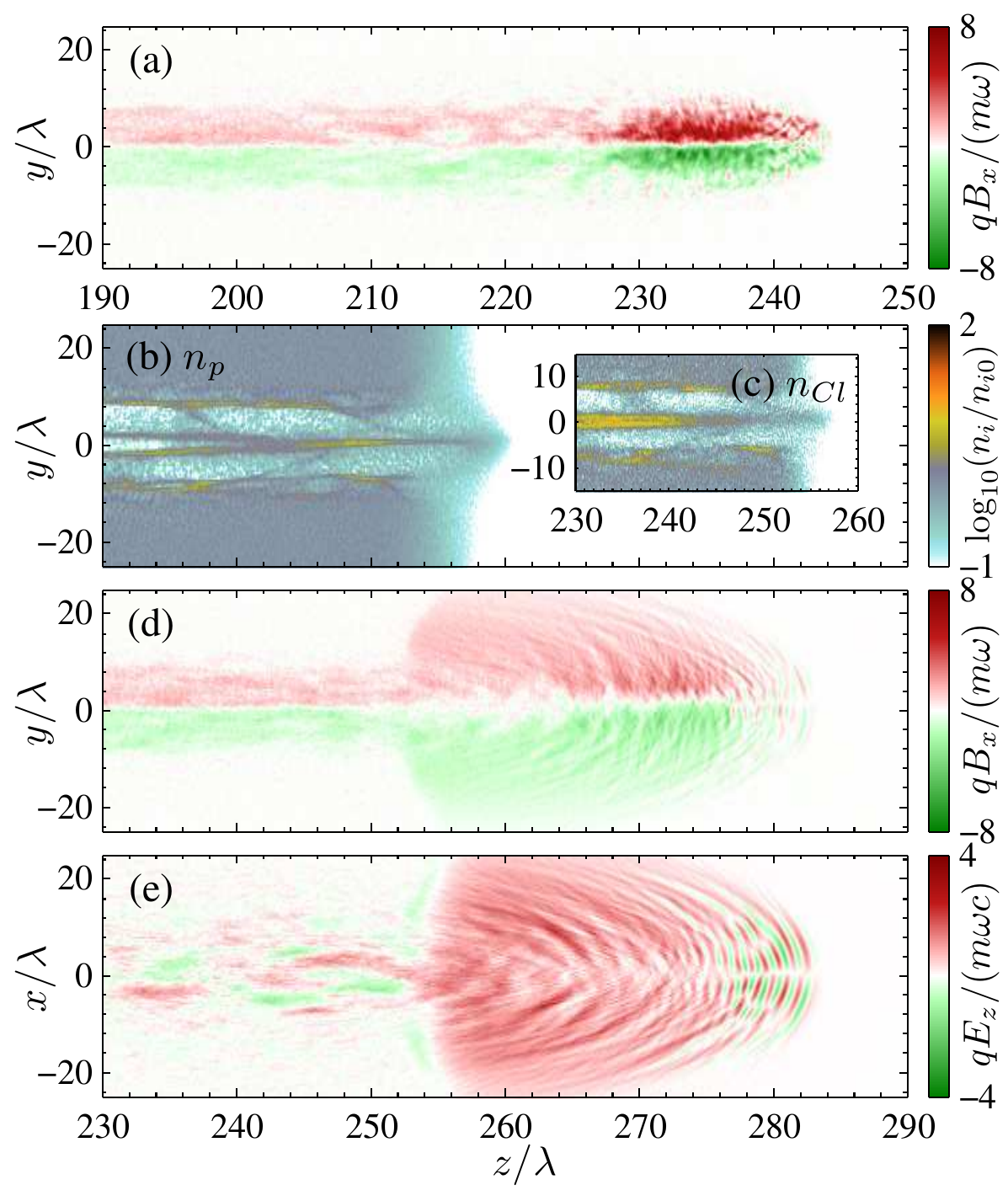}
    \caption{Snapshots of: (a) azimuthal magnetic field at $t=250\lambda/c$, and (b) proton density, (c) chlorine nuclei density, (d)  azimuthal magnetic field and (e) longitudinal electric field at $t=290\lambda/c$.}
    \label{fig:3}
\end{figure}

Acceleration of protons and spin precession take place within the plasma channel driven by the intense laser. A plasma channel is formed by ponderomotive expulsion of charge lying within the laser's path, as a relativistic self-focusing effect in high power laser plasma interactions when the laser power exceeds the critical value $P_c=17 n_c/n_0$ GW.  When a laser pulse with even higher power propagates in an underdense plasma, $P_L=21.49a_0^2w_0^2/\lambda^2~\mathrm{GW}\gg P_c$, an additional central electron filament is enclosed in the evacuated channel~\cite{Popov:2010:PhysRevLett.105.195002, Ji:2014:PhysRevLett.112.145003} The displacement of electrons introduces a radial electric field $E_r$ in the plasma channel, as well as an azimuthal magnetic field around the channel axis $B_\phi$. The magnetic field $B_\phi$ is represented in Fig. \ref{fig:3}(a) by $B_x$ in the $z-y$ plane. Effected by the space charge field $E_r$, ions also move into the filament along the channel axis~\cite{Sylla:2012:PhysRevLett.108.115003}.
Due to the small charge-to-mass ratio of ions, they converge towards the center  on a longer time scale as compared to the electrons. More specifically, the electron filament appears inside the laser fields, while the proton filament forms behind the laser pulse but in front of the chlorine nuclei filament. Figures \ref{fig:3}(b) and (c) show the protn filament is found around  $240\lesssim x/\lambda \lesssim 260$, and the filament of chlorine nuclei lies around $x\lesssim 240\lambda$. Because of the different responds of particles, longitudinal and radial space-charge fields are formed within the plasma channel. As a results, the protons are accelerated in forward and transverse directions under the actions of the electrons in front and the chlorine nuclei behind.
The advantage of the admixture of chlorine is analogous to the case investigated in Ref.~\cite{Shen:2007:PhysRevE.76.055402}.  

In addition to the first-stage acceleration in the plasma
channel, we find a second effect at the rear end of the gas jet, which enhances the proton energies even more significantly. When the electrons driven by the laser pulse pass though the rear boundary, the azimuthal magnetic fields expands into the vacuum under large angles. Transverse expansion of the magnetic field generates a strong longitudinal field according to the Faraday's law, as shown in Fig. \ref{fig:3}(d) and (e). The focused protons in the filament near the rear boundary are then accelerated strongly. This secondary acceleration is known as magnetic vortex acceleration~\cite{Nakamura:2010:PhysRevLett.105.135002, kawata:HPLSE:2014,Park:2019:POP}.

\begin{figure}[t]
    \includegraphics[width=8.5cm]{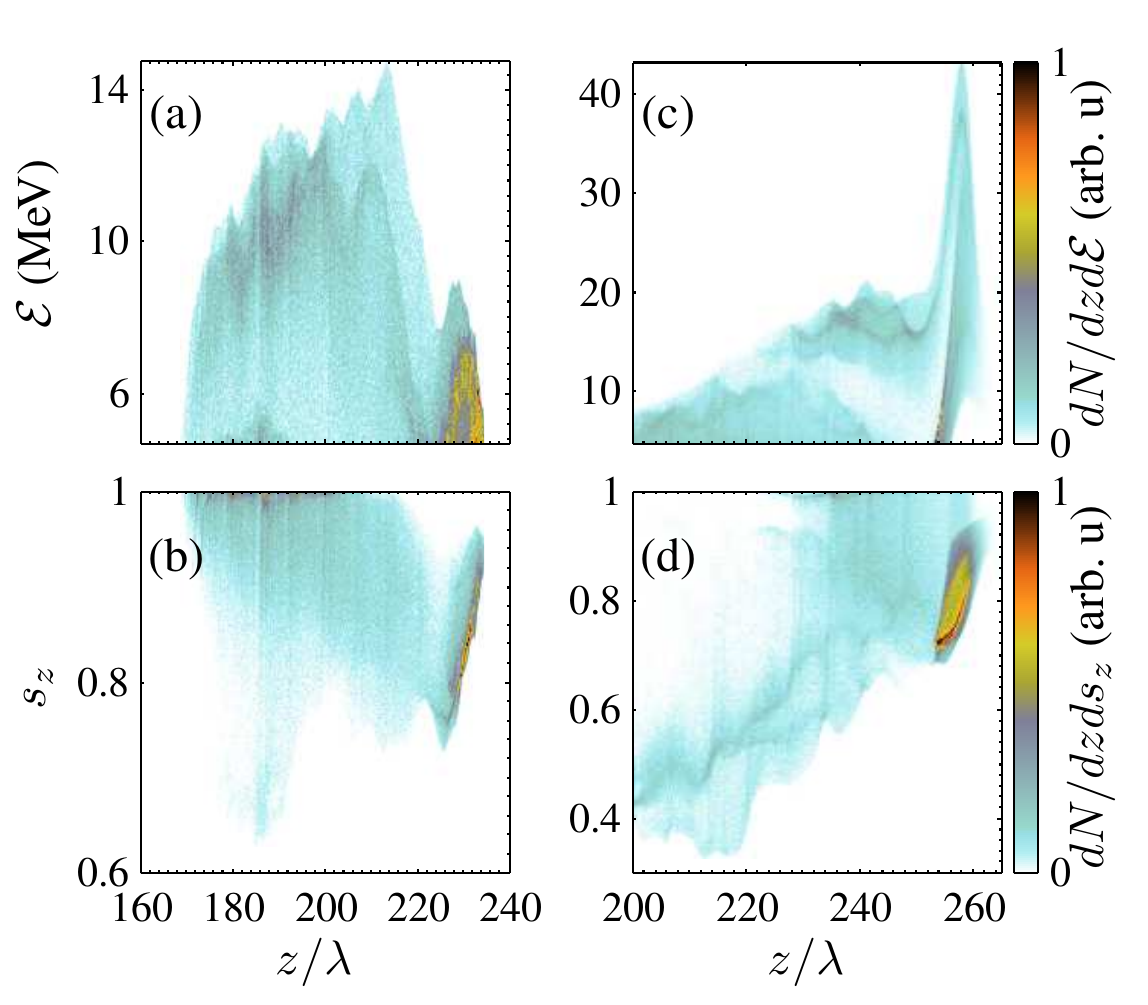}
    \caption{Density plot of energetic protons (with $\mathcal{E}\geqslant5$ MeV) in (a) (c) phase space and (b) (d) space with spin component $s_z$ and $z$-axis. (a), (b) and (c), (d) corresponds to density plots at $t=250\lambda/c$ and $t=290\lambda/c$, respectively.}
    \label{fig:4}
\end{figure}

Figures \ref{fig:4}(a) and (b) present effects of the proton acceleration in the plasma channel (with $z<z_2$), and the spin distribution of energetic protons, respectively. It indicates at $t=250 \lambda/c$ many energetic protons start to converge to the front of the proton filament around $z_f=230\lambda$. The energy of these \textit{front protons} is less than 10 MeV. In contrast to that, the \textit{tail protons} which converge to the filament before ($t<250\lambda/c$) are accelerated further to energies beyond 10 MeV but sit far behind the driven laser with $z<z_f$. The spin precession for these front protons and tail protons play different roles as will be discussed below. The PIC simulations have provided clear evidences for the strong magnetic field in the laser driven plasma channel as shown in Figs. \ref{fig:3}(a) and (c). Due to the cylindrical symmetry of the azimuthal magnetic field in the plasma channel, the transverse spin components, $s_x$ and $s_y$, spread symmetrically. This concurs with the spin distribution on the Bloch sphere shown in Fig. \ref{fig:2}(b). The polarization of the collected proton sample is thus determined by the longitudinal spin component $P=\left<s_z\right>$. Note that depolarization via direct precession of protons is donimant in laser accelerators and the depolarization rate of protons in plasmas via binary collisions $\sim 10^{-22} n_\mathrm{cm3}\sqrt{T_\mathrm{eV}}~\mathrm{s^{-1}}$ is negligibly small because of the applied low densities and low temperatures~\cite{Kulsrud:1986:Nuclear-Fusion}, where $T_\mathrm{eV}$ is the temperature in units of electronvote and $n_\mathrm{cm3}$ is the plasma density in units of $\mathrm{cm^{-3}}$. 

Spin precession is calculated in the PIC simulations with the Thomas-Bargmann-Michel-Telegdi equation~\cite{Thomas:1926:Motion_of, Bargmann:Telegdi:1959:Precession_of}
\begin{align}
\frac{d\vec{s}}{dt}=\frac{q}{M}\vec{s}\times\left[\left(G+\frac{1}{\gamma}\right)\vec{B} - \frac{G\gamma \vec{v}\cdot\vec{B}}{\gamma+1}\frac{\vec{v}}{c^2} \right. \nonumber \\  \left.
- \left(G+\frac{1}{\gamma+1}\right)\frac{\vec{v}}{c^2} \times\vec{E}\right] ,\quad
\end{align}
where $G\approx 1.79$ is the anomalous magnetic moment of the proton and $M$ and is the proton mass. Simulations as shown in Figs. \ref{fig:4}(a) and (c) indicate $\gamma-1\sim 10^{-2}$ and $\left| \vec v\right|/c \sim 10^{-1}$ for  accelerated protons. The spin of protons in the filament can be roughly estimated via \mbox{$s_z \approx 1-\int_0^\tau dt (q/M) (G+1)B_\phi$}, with $\tau$ the time duration for proton to move to the axis under the radial electric field $E_r$. It explains a significant spin precession in the simulations. First, due to the small proton velocity, the precession time $\tau$ in the magnetic fields is long; second, the anomalous magnetic moment $G$ is larger than that of electrons; third, spin precession of a proton is directly affected by its trajectory and the fields structure, where the dominant azimuthal magnetic field  $B_\phi \sim 8 m\omega/q$ in the plasma channel is strong. 

When the protons are focused to the axis, the longitudinal spin component of front protons around $z=z_f$ decreases to $s_z \sim 80\%$ as shown in Fig. \ref{fig:4}(b). This indicates that the magnetic field is strong enough to significantly depolarize the accelerated beam protons. In the channel driven by lasers with $a_0=25$, about $80~\%$ of the polarization is retained. For tail protons with further acceleration in the channel $z<z_f$ in Fig. \ref{fig:4}(a), they conserve their momentum and continue to travel radially away after they cross the laser axis, which results in a change of the direction of $B_\phi$. Figure \ref{fig:4}(b) shows larger $s_z$ of these protons indicating they precess back to their initial direction along $z$.

At the rear boundary of the gas jet, protons experience the second stage acceleration. When the front of the proton filament passes $z_2$ immediately after the electrons and enters  the region with strong longitudinal electric field [see Figs. \ref{fig:3}(c) and (d)], corresponding protons are accelerated intensely to  40 MeV [see Fig. \ref{fig:4}(c)]. Meanwhile, the spin vector of front protons precesses roughly at the same rate ($\sim 80\%$) compared to the front protons in the plasma channel, see Figs. \ref{fig:4}(b) and (d). Differently, the more energetic tail protons in the channel leave too far behind the electrons. When they pass  $z_2$, the magnetic vortex induced electric field becomes too weak to accelerate protons for its limited lifetime. Figure \ref{fig:2}(a) indicates that only the front protons in the filament are accelerated to an energy $>20$ MeV. In other words, the tail protons neither contribute to the final accelerated proton beam nor to the beam polarization.

\begin{figure}[t]
    \includegraphics[width=8.5cm]{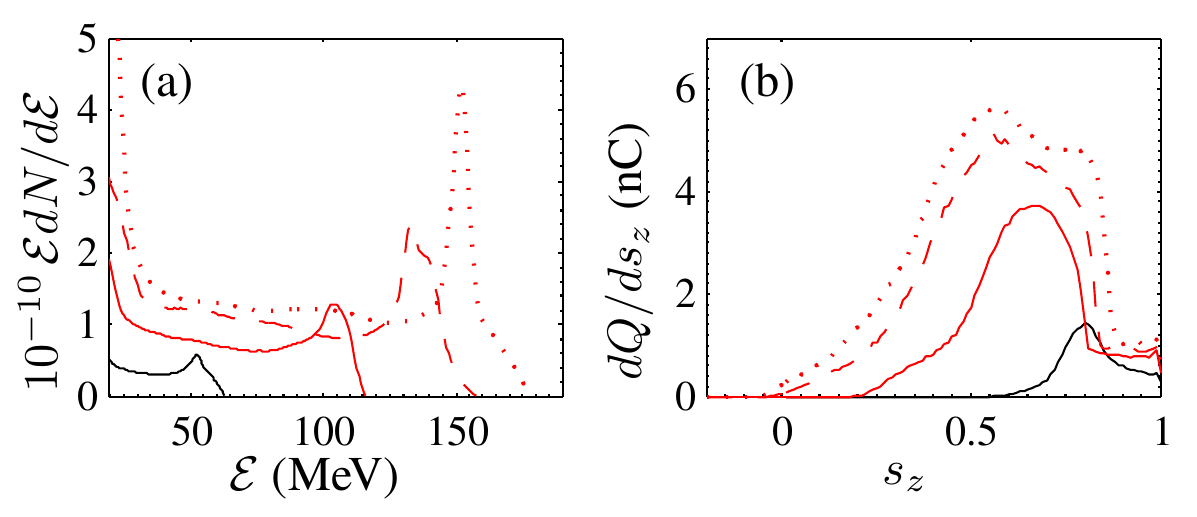}
    \caption{Snapshots at $t=330\lambda/c$ of: (a) Energy spectra and (b) spin distribution of accelerated protonswith $\mathcal{E}\geqslant 20$ MeV in cases with $a_0=25$, $50$, $75$ and $100$, which are presented with solid-black, solid-red, dashed-red and dotted-red curves, respectively.}
    \label{fig:5}
\end{figure}

\begin{table}
	\caption{\label{beams} Polarized beams accelerated by PW lasers. Results from Fig. \ref{fig:5} are compared  and the parameters used are given. Here, $P_L$ is the laser power, $\mathcal {E}_p$ the peak proton energy, $Q$ the total charge, $P$ the beam polarization.}
	\begin{ruledtabular}
		\begin{tabular}{lcccr}
			$a_0$ & $P_L$ [PW] & $\mathcal{E}_p$ [MeV] & $Q$ [nC] & $P$ [\%] \\
			\hline
			25 & 1.34 & 53 & 0.26 & 82 \\
			50 & 5.37 & 105 & 1.3 & 65 \\
			75 & 12.1 & 133 & 2.4 & 57 \\
			100 & 21.5 & 152 & 3.1 & 56 \\
		\end{tabular}
	\end{ruledtabular}
\end{table}

After the second-stage acceleration at the rear boundary of the gas jet, electromagnetic fields around the accelerated beam quickly decay. The magnetic field as well as the electric field around the propagating axis become too weak to affect the protons after $t=330\lambda/c$. Thus both energy and polarization of the proton beam remains constant. Properties of the final accelerated proton beams with $\mathcal{E}\geqslant 20$ MeV are presented in Fig. \ref{fig:2} and solid-black lines in Fig. \ref{fig:5}. Applying laser systems with a power of 1.3 PW, the protons are accelerated to tens of MeV with a peak at 53 MeV, while polarization of $P=82~\%$ is retained. The corresponding proton current shown in Fig. \ref{fig:2}(a) is of the order of 10 kiloampere. Simulations with lower target density ($n_p\sim 10^{18} \mathrm{cm^{-3}}$) results in lower charge of the accelerated polarized proton beam (not shown). The accelerated energy and beam charge, as listed in Table \ref{beams}, increase with the driving laser intensity. As a polarized proton source, one may not boost the proton energy by enhancing the plasma density as other magnetic vortex acceleration \cite{Nakamura:2010:PhysRevLett.105.135002}. Therefore the beam energy does not increase with the laser power linearly. Figure \ref{fig:5}(b) implies this non-linear dependence on the laser power is also applies to the proton depolarization. Low intense laser pulses or the rising edges of the main pulse with much lower power in a real experiment are not intense enough to accelerate and depolarize the proton beams from a gas jet. While totally polarized HCl gases are possibly produced experimentally~\cite{Rakitzis:Science:2003:1936, Sofikitis:2017PhysRevLett.118.233401},  the beam polarization can be reduced to $\eta_0 P$ when $\eta_0<1$. A systematic study of other experimental parameters, like the inhomogeneity of initial polarization $\eta_0\left(x,y,z\right)$, pulse intensity and duration, or target density and length, on the proton energy distribution, flux and polarization is under way and will be subject of a forthcoming publication. More importantly, as shown in Fig. \ref{fig:5}(b) and Table \ref{beams}, more than 50 \% of the proton beam polarization driven by PW lasers can be sustained. Although the beam quality need to be further improved, effective detection in high energy experiments may benefit from the high luminosity of the laser driven proton beams.

\begin{acknowledgments}
We would like to thank Z.Y.Li for helpful discussions. This work was supported by the Ministry of Science and
Technology of the People’s Republic of China (Grant Nos.
2018YFA0404803, 2016YFA0401102), the National Natural
Science Foundation of China (Grant Nos. 11674339,
11922515, 11935008), the Strategic Priority Research Program
of the Chinese Academy of Sciences (Grant No.
XDB16) and Innovation Program of Shanghai Municipal
Education Commission. The work of M.B. and A.H. has been carried out in the framework of the {\it Ju}SPARC (J{\" u}lich Short-Pulse Particle and Radiation Center) and has been supported by the
ATHENA (Accelerator Technology HElmholtz iNfrAstructure) consortium.
\end{acknowledgments}

\nocite{*}

\bibliography{wref}

\begin{thebibliography}{51}%
\makeatletter
\providecommand \@ifxundefined [1]{%
 \@ifx{#1\undefined}
}%
\providecommand \@ifnum [1]{%
 \ifnum #1\expandafter \@firstoftwo
 \else \expandafter \@secondoftwo
 \fi
}%
\providecommand \@ifx [1]{%
 \ifx #1\expandafter \@firstoftwo
 \else \expandafter \@secondoftwo
 \fi
}%
\providecommand \natexlab [1]{#1}%
\providecommand \enquote  [1]{``#1''}%
\providecommand \bibnamefont  [1]{#1}%
\providecommand \bibfnamefont [1]{#1}%
\providecommand \citenamefont [1]{#1}%
\providecommand \href@noop [0]{\@secondoftwo}%
\providecommand \href [0]{\begingroup \@sanitize@url \@href}%
\providecommand \@href[1]{\@@startlink{#1}\@@href}%
\providecommand \@@href[1]{\endgroup#1\@@endlink}%
\providecommand \@sanitize@url [0]{\catcode `\\12\catcode `\$12\catcode
  `\&12\catcode `\#12\catcode `\^12\catcode `\_12\catcode `\%12\relax}%
\providecommand \@@startlink[1]{}%
\providecommand \@@endlink[0]{}%
\providecommand \url  [0]{\begingroup\@sanitize@url \@url }%
\providecommand \@url [1]{\endgroup\@href {#1}{\urlprefix }}%
\providecommand \urlprefix  [0]{URL }%
\providecommand \Eprint [0]{\href }%
\providecommand \doibase [0]{https://doi.org/}%
\providecommand \selectlanguage [0]{\@gobble}%
\providecommand \bibinfo  [0]{\@secondoftwo}%
\providecommand \bibfield  [0]{\@secondoftwo}%
\providecommand \translation [1]{[#1]}%
\providecommand \BibitemOpen [0]{}%
\providecommand \bibitemStop [0]{}%
\providecommand \bibitemNoStop [0]{.\EOS\space}%
\providecommand \EOS [0]{\spacefactor3000\relax}%
\providecommand \BibitemShut  [1]{\csname bibitem#1\endcsname}%
\let\auto@bib@innerbib\@empty
\bibitem [{\citenamefont {Gentile}\ \emph {et~al.}(2017)\citenamefont
  {Gentile}, \citenamefont {Nacher}, \citenamefont {Saam},\ and\ \citenamefont
  {Walker}}]{Gentile:2017:RevModPhys.89.045004}%
  \BibitemOpen
  \bibfield  {author} {\bibinfo {author} {\bibfnamefont {T.~R.}\ \bibnamefont
  {Gentile}}, \bibinfo {author} {\bibfnamefont {P.~J.}\ \bibnamefont {Nacher}},
  \bibinfo {author} {\bibfnamefont {B.}~\bibnamefont {Saam}},\ and\ \bibinfo
  {author} {\bibfnamefont {T.~G.}\ \bibnamefont {Walker}},\ }\bibfield  {title}
  {\bibinfo {title} {Optically polarized $^{3}\mathrm{He}$},\ }\href
  {https://doi.org/10.1103/RevModPhys.89.045004} {\bibfield  {journal}
  {\bibinfo  {journal} {Rev. Mod. Phys.}\ }\textbf {\bibinfo {volume} {89}},\
  \bibinfo {pages} {045004} (\bibinfo {year} {2017})}\BibitemShut {NoStop}%
\bibitem [{\citenamefont {Safronova}\ \emph {et~al.}(2018)\citenamefont
  {Safronova}, \citenamefont {Budker}, \citenamefont {DeMille}, \citenamefont
  {Kimball}, \citenamefont {Derevianko},\ and\ \citenamefont
  {Clark}}]{Safronova:2018:RevModPhys.90.025008}%
  \BibitemOpen
  \bibfield  {author} {\bibinfo {author} {\bibfnamefont {M.~S.}\ \bibnamefont
  {Safronova}}, \bibinfo {author} {\bibfnamefont {D.}~\bibnamefont {Budker}},
  \bibinfo {author} {\bibfnamefont {D.}~\bibnamefont {DeMille}}, \bibinfo
  {author} {\bibfnamefont {D.~F.~J.}\ \bibnamefont {Kimball}}, \bibinfo
  {author} {\bibfnamefont {A.}~\bibnamefont {Derevianko}},\ and\ \bibinfo
  {author} {\bibfnamefont {C.~W.}\ \bibnamefont {Clark}},\ }\bibfield  {title}
  {\bibinfo {title} {Search for new physics with atoms and molecules},\ }\href
  {https://doi.org/10.1103/RevModPhys.90.025008} {\bibfield  {journal}
  {\bibinfo  {journal} {Rev. Mod. Phys.}\ }\textbf {\bibinfo {volume} {90}},\
  \bibinfo {pages} {025008} (\bibinfo {year} {2018})}\BibitemShut {NoStop}%
\bibitem [{\citenamefont {Moortgat-Pick}\ \emph {et~al.}(2008)\citenamefont
  {Moortgat-Pick} \emph {et~al.}}]{MoortgatPick:2005cw}%
  \BibitemOpen
  \bibfield  {author} {\bibinfo {author} {\bibfnamefont {G.}~\bibnamefont
  {Moortgat-Pick}} \emph {et~al.},\ }\bibfield  {title} {\bibinfo {title} {{The
  Role of polarized positrons and electrons in revealing fundamental
  interactions at the linear collider}},\ }\href
  {https://doi.org/10.1016/j.physrep.2007.12.003} {\bibfield  {journal}
  {\bibinfo  {journal} {Phys. Rept.}\ }\textbf {\bibinfo {volume} {460}},\
  \bibinfo {pages} {131} (\bibinfo {year} {2008})},\ \Eprint
  {https://arxiv.org/abs/hep-ph/0507011} {arXiv:hep-ph/0507011} \BibitemShut
  {NoStop}%
\bibitem [{\citenamefont {Mane}\ \emph {et~al.}(2005)\citenamefont {Mane},
  \citenamefont {Shatunov},\ and\ \citenamefont {Yokoya}}]{Mane:RPP:2005}%
  \BibitemOpen
  \bibfield  {author} {\bibinfo {author} {\bibfnamefont {S.~R.}\ \bibnamefont
  {Mane}}, \bibinfo {author} {\bibfnamefont {Y.~M.}\ \bibnamefont {Shatunov}},\
  and\ \bibinfo {author} {\bibfnamefont {K.}~\bibnamefont {Yokoya}},\
  }\bibfield  {title} {\bibinfo {title} {Spin-polarized charged particle beams
  in high-energy accelerators},\ }\href
  {http://stacks.iop.org/0034-4885/68/i=9/a=R01} {\bibfield  {journal}
  {\bibinfo  {journal} {Rep. Prog. Phys.}\ }\textbf {\bibinfo {volume} {68}},\
  \bibinfo {pages} {1997} (\bibinfo {year} {2005})}\BibitemShut {NoStop}%
\bibitem [{\citenamefont {Adlarson}\ \emph {et~al.}(2014)\citenamefont
  {Adlarson}, \citenamefont {Augustyniak}, \citenamefont {Bardan},
  \citenamefont {Bashkanov}, \citenamefont {Bergmann}, \citenamefont
  {Ber\l{}owski}, \citenamefont {Bhatt}, \citenamefont {B\"uscher},
  \citenamefont {Cal\'en}, \citenamefont {Ciepa\l{}} \emph
  {et~al.}}]{Adlarson:2014:PhysRevLett.112.202301}%
  \BibitemOpen
  \bibfield  {author} {\bibinfo {author} {\bibfnamefont {P.}~\bibnamefont
  {Adlarson}}, \bibinfo {author} {\bibfnamefont {W.}~\bibnamefont
  {Augustyniak}}, \bibinfo {author} {\bibfnamefont {W.}~\bibnamefont {Bardan}},
  \bibinfo {author} {\bibfnamefont {M.}~\bibnamefont {Bashkanov}}, \bibinfo
  {author} {\bibfnamefont {F.~S.}\ \bibnamefont {Bergmann}}, \bibinfo {author}
  {\bibfnamefont {M.}~\bibnamefont {Ber\l{}owski}}, \bibinfo {author}
  {\bibfnamefont {H.}~\bibnamefont {Bhatt}}, \bibinfo {author} {\bibfnamefont
  {M.}~\bibnamefont {B\"uscher}}, \bibinfo {author} {\bibfnamefont
  {H.}~\bibnamefont {Cal\'en}}, \bibinfo {author} {\bibfnamefont
  {I.}~\bibnamefont {Ciepa\l{}}}, \emph {et~al.} (\bibinfo {collaboration}
  {WASA-at-COSY Collaboration and SAID Data Analysis Center}),\ }\bibfield
  {title} {\bibinfo {title} {Evidence for a new resonance from polarized
  neutron-proton scattering},\ }\href
  {https://doi.org/10.1103/PhysRevLett.112.202301} {\bibfield  {journal}
  {\bibinfo  {journal} {Phys. Rev. Lett.}\ }\textbf {\bibinfo {volume} {112}},\
  \bibinfo {pages} {202301} (\bibinfo {year} {2014})}\BibitemShut {NoStop}%
\bibitem [{\citenamefont {Adamczyk}\ \emph {et~al.}(2016)\citenamefont
  {Adamczyk}, \citenamefont {Adkins}, \citenamefont {Agakishiev}, \citenamefont
  {Aggarwal}, \citenamefont {Ahammed}, \citenamefont {Alekseev}, \citenamefont
  {Aparin}, \citenamefont {Arkhipkin}, \citenamefont {Aschenauer},
  \citenamefont {Attri} \emph {et~al.}}]{Adamczyk:2016:PhysRevLett.116.132301}%
  \BibitemOpen
  \bibfield  {author} {\bibinfo {author} {\bibfnamefont {L.}~\bibnamefont
  {Adamczyk}}, \bibinfo {author} {\bibfnamefont {J.~K.}\ \bibnamefont
  {Adkins}}, \bibinfo {author} {\bibfnamefont {G.}~\bibnamefont {Agakishiev}},
  \bibinfo {author} {\bibfnamefont {M.~M.}\ \bibnamefont {Aggarwal}}, \bibinfo
  {author} {\bibfnamefont {Z.}~\bibnamefont {Ahammed}}, \bibinfo {author}
  {\bibfnamefont {I.}~\bibnamefont {Alekseev}}, \bibinfo {author}
  {\bibfnamefont {A.}~\bibnamefont {Aparin}}, \bibinfo {author} {\bibfnamefont
  {D.}~\bibnamefont {Arkhipkin}}, \bibinfo {author} {\bibfnamefont {E.~C.}\
  \bibnamefont {Aschenauer}}, \bibinfo {author} {\bibfnamefont
  {A.}~\bibnamefont {Attri}}, \emph {et~al.} (\bibinfo {collaboration} {STAR
  Collaboration}),\ }\bibfield  {title} {\bibinfo {title} {Measurement of the
  transverse single-spin asymmetry in
  ${p}^{\ensuremath{\uparrow}}+p\ensuremath{\rightarrow}{W}^{\ifmmode\pm\else\textpm\fi{}}/{Z}^{0}$
  at rhic},\ }\href {https://doi.org/10.1103/PhysRevLett.116.132301} {\bibfield
   {journal} {\bibinfo  {journal} {Phys. Rev. Lett.}\ }\textbf {\bibinfo
  {volume} {116}},\ \bibinfo {pages} {132301} (\bibinfo {year}
  {2016})}\BibitemShut {NoStop}%
\bibitem [{\citenamefont {Aschenauer}\ \emph {et~al.}(2019)\citenamefont
  {Aschenauer}, \citenamefont {Fazio}, \citenamefont {Lee}, \citenamefont
  {M\"{a}ntysaari}, \citenamefont {Page}, \citenamefont {Schenke},
  \citenamefont {Ullrich}, \citenamefont {Venugopalan},\ and\ \citenamefont
  {Zurita}}]{Aschenauer:RPP:2019}%
  \BibitemOpen
  \bibfield  {author} {\bibinfo {author} {\bibfnamefont {E.~C.}\ \bibnamefont
  {Aschenauer}}, \bibinfo {author} {\bibfnamefont {S.}~\bibnamefont {Fazio}},
  \bibinfo {author} {\bibfnamefont {J.~H.}\ \bibnamefont {Lee}}, \bibinfo
  {author} {\bibfnamefont {H.}~\bibnamefont {M\"{a}ntysaari}}, \bibinfo
  {author} {\bibfnamefont {B.~S.}\ \bibnamefont {Page}}, \bibinfo {author}
  {\bibfnamefont {B.}~\bibnamefont {Schenke}}, \bibinfo {author} {\bibfnamefont
  {T.}~\bibnamefont {Ullrich}}, \bibinfo {author} {\bibfnamefont
  {R.}~\bibnamefont {Venugopalan}},\ and\ \bibinfo {author} {\bibfnamefont
  {P.}~\bibnamefont {Zurita}},\ }\bibfield  {title} {\bibinfo {title} {The
  electron–ion collider: assessing the energy dependence of key
  measurements},\ }\href {https://iopscience.iop.org/0034-4885/82/2/024301/}
  {\bibfield  {journal} {\bibinfo  {journal} {Rep. Prog. Phys.}\ }\textbf
  {\bibinfo {volume} {82}},\ \bibinfo {pages} {024301} (\bibinfo {year}
  {2019})}\BibitemShut {NoStop}%
\bibitem [{\citenamefont {de~Florian}\ \emph {et~al.}(2014)\citenamefont
  {de~Florian}, \citenamefont {Sassot}, \citenamefont {Stratmann},\ and\
  \citenamefont {Vogelsang}}]{Florian:2014:PhysRevLett.113.012001}%
  \BibitemOpen
  \bibfield  {author} {\bibinfo {author} {\bibfnamefont {D.}~\bibnamefont
  {de~Florian}}, \bibinfo {author} {\bibfnamefont {R.}~\bibnamefont {Sassot}},
  \bibinfo {author} {\bibfnamefont {M.}~\bibnamefont {Stratmann}},\ and\
  \bibinfo {author} {\bibfnamefont {W.}~\bibnamefont {Vogelsang}},\ }\bibfield
  {title} {\bibinfo {title} {Evidence for polarization of gluons in the
  proton},\ }\href {https://doi.org/10.1103/PhysRevLett.113.012001} {\bibfield
  {journal} {\bibinfo  {journal} {Phys. Rev. Lett.}\ }\textbf {\bibinfo
  {volume} {113}},\ \bibinfo {pages} {012001} (\bibinfo {year}
  {2014})}\BibitemShut {NoStop}%
\bibitem [{\citenamefont {Alexandrou}\ \emph {et~al.}(2017)\citenamefont
  {Alexandrou}, \citenamefont {Constantinou}, \citenamefont {Hadjiyiannakou},
  \citenamefont {Jansen}, \citenamefont {Kallidonis}, \citenamefont {Koutsou},
  \citenamefont {Avil\'es-Casco},\ and\ \citenamefont
  {Wiese}}]{Alexandrou:2017:PhysRevLett.119.142002}%
  \BibitemOpen
  \bibfield  {author} {\bibinfo {author} {\bibfnamefont {C.}~\bibnamefont
  {Alexandrou}}, \bibinfo {author} {\bibfnamefont {M.}~\bibnamefont
  {Constantinou}}, \bibinfo {author} {\bibfnamefont {K.}~\bibnamefont
  {Hadjiyiannakou}}, \bibinfo {author} {\bibfnamefont {K.}~\bibnamefont
  {Jansen}}, \bibinfo {author} {\bibfnamefont {C.}~\bibnamefont {Kallidonis}},
  \bibinfo {author} {\bibfnamefont {G.}~\bibnamefont {Koutsou}}, \bibinfo
  {author} {\bibfnamefont {A.~V.}\ \bibnamefont {Avil\'es-Casco}},\ and\
  \bibinfo {author} {\bibfnamefont {C.}~\bibnamefont {Wiese}},\ }\bibfield
  {title} {\bibinfo {title} {Nucleon spin and momentum decomposition using
  lattice qcd simulations},\ }\href
  {https://doi.org/10.1103/PhysRevLett.119.142002} {\bibfield  {journal}
  {\bibinfo  {journal} {Phys. Rev. Lett.}\ }\textbf {\bibinfo {volume} {119}},\
  \bibinfo {pages} {142002} (\bibinfo {year} {2017})}\BibitemShut {NoStop}%
\bibitem [{\citenamefont {Androic}\ \emph {et~al.}(2018)\citenamefont {Androic}
  \emph {et~al.}}]{Androic2018}%
  \BibitemOpen
  \bibfield  {author} {\bibinfo {author} {\bibfnamefont {D.}~\bibnamefont
  {Androic}} \emph {et~al.} (\bibinfo {collaboration} {The Jefferson Lab Qweak
  Collaboration}),\ }\bibfield  {title} {\bibinfo {title} {Precision
  measurement of the weak charge of the proton},\ }\href
  {https://doi.org/10.1038/s41586-018-0096-0} {\bibfield  {journal} {\bibinfo
  {journal} {Nature}\ }\textbf {\bibinfo {volume} {557}},\ \bibinfo {pages}
  {207} (\bibinfo {year} {2018})}\BibitemShut {NoStop}%
\bibitem [{\citenamefont {Accardi}\ \emph {et~al.}(2016)\citenamefont
  {Accardi}, \citenamefont {Albacete}, \citenamefont {Anselmino}, \citenamefont
  {Armesto}, \citenamefont {Aschenauer}, \citenamefont {Bacchetta},
  \citenamefont {Boer}, \citenamefont {Brooks}, \citenamefont {Burton},
  \citenamefont {Chang} \emph {et~al.}}]{Accardi2016}%
  \BibitemOpen
  \bibfield  {author} {\bibinfo {author} {\bibfnamefont {A.}~\bibnamefont
  {Accardi}}, \bibinfo {author} {\bibfnamefont {J.~L.}\ \bibnamefont
  {Albacete}}, \bibinfo {author} {\bibfnamefont {M.}~\bibnamefont {Anselmino}},
  \bibinfo {author} {\bibfnamefont {N.}~\bibnamefont {Armesto}}, \bibinfo
  {author} {\bibfnamefont {E.~C.}\ \bibnamefont {Aschenauer}}, \bibinfo
  {author} {\bibfnamefont {A.}~\bibnamefont {Bacchetta}}, \bibinfo {author}
  {\bibfnamefont {D.}~\bibnamefont {Boer}}, \bibinfo {author} {\bibfnamefont
  {W.~K.}\ \bibnamefont {Brooks}}, \bibinfo {author} {\bibfnamefont
  {T.}~\bibnamefont {Burton}}, \bibinfo {author} {\bibfnamefont {N.~B.}\
  \bibnamefont {Chang}}, \emph {et~al.},\ }\bibfield  {title} {\bibinfo {title}
  {Electron-ion collider: The next qcd frontier},\ }\href
  {https://doi.org/10.1140/epja/i2016-16268-9} {\bibfield  {journal} {\bibinfo
  {journal} {Eur. Phys. J. A}\ }\textbf {\bibinfo {volume} {52}},\ \bibinfo
  {pages} {268} (\bibinfo {year} {2016})}\BibitemShut {NoStop}%
\bibitem [{\citenamefont {Yanovsky}\ \emph {et~al.}(2008)\citenamefont
  {Yanovsky}, \citenamefont {Chvykov}, \citenamefont {Kalinchenko},
  \citenamefont {Rousseau}, \citenamefont {Planchon}, \citenamefont {Matsuoka},
  \citenamefont {Maksimchuk}, \citenamefont {Nees}, \citenamefont {Cheriaux},
  \citenamefont {Mourou} \emph {et~al.}}]{Yanovsky:OE:2008}%
  \BibitemOpen
  \bibfield  {author} {\bibinfo {author} {\bibfnamefont {V.}~\bibnamefont
  {Yanovsky}}, \bibinfo {author} {\bibfnamefont {V.}~\bibnamefont {Chvykov}},
  \bibinfo {author} {\bibfnamefont {G.}~\bibnamefont {Kalinchenko}}, \bibinfo
  {author} {\bibfnamefont {P.}~\bibnamefont {Rousseau}}, \bibinfo {author}
  {\bibfnamefont {T.}~\bibnamefont {Planchon}}, \bibinfo {author}
  {\bibfnamefont {T.}~\bibnamefont {Matsuoka}}, \bibinfo {author}
  {\bibfnamefont {A.}~\bibnamefont {Maksimchuk}}, \bibinfo {author}
  {\bibfnamefont {J.}~\bibnamefont {Nees}}, \bibinfo {author} {\bibfnamefont
  {G.}~\bibnamefont {Cheriaux}}, \bibinfo {author} {\bibfnamefont
  {G.}~\bibnamefont {Mourou}}, \emph {et~al.},\ }\bibfield  {title} {\bibinfo
  {title} {Ultra-high intensity- {300-TW laser at 0.1 Hz} repetition rate.},\
  }\href {https://doi.org/10.1364/OE.16.002109} {\bibfield  {journal} {\bibinfo
   {journal} {Opt. Express}\ }\textbf {\bibinfo {volume} {16}},\ \bibinfo
  {pages} {2109} (\bibinfo {year} {2008})}\BibitemShut {NoStop}%
\bibitem [{\citenamefont {Guo}\ \emph {et~al.}(2018)\citenamefont {Guo},
  \citenamefont {Yu}, \citenamefont {Wang}, \citenamefont {Wang}, \citenamefont
  {Liu}, \citenamefont {Gan}, \citenamefont {Li}, \citenamefont {Leng},
  \citenamefont {Liang},\ and\ \citenamefont {Li}}]{Guo:OE:2018}%
  \BibitemOpen
  \bibfield  {author} {\bibinfo {author} {\bibfnamefont {Z.}~\bibnamefont
  {Guo}}, \bibinfo {author} {\bibfnamefont {L.}~\bibnamefont {Yu}}, \bibinfo
  {author} {\bibfnamefont {J.}~\bibnamefont {Wang}}, \bibinfo {author}
  {\bibfnamefont {C.}~\bibnamefont {Wang}}, \bibinfo {author} {\bibfnamefont
  {Y.}~\bibnamefont {Liu}}, \bibinfo {author} {\bibfnamefont {Z.}~\bibnamefont
  {Gan}}, \bibinfo {author} {\bibfnamefont {W.}~\bibnamefont {Li}}, \bibinfo
  {author} {\bibfnamefont {Y.}~\bibnamefont {Leng}}, \bibinfo {author}
  {\bibfnamefont {X.}~\bibnamefont {Liang}},\ and\ \bibinfo {author}
  {\bibfnamefont {R.}~\bibnamefont {Li}},\ }\bibfield  {title} {\bibinfo
  {title} {Improvement of the focusing ability by double deformable mirrors for
  {10-PW-level Ti:} sapphire chirped pulse amplification laser system},\ }\href
  {https://doi.org/10.1364/OE.26.026776} {\bibfield  {journal} {\bibinfo
  {journal} {Opt. Express}\ }\textbf {\bibinfo {volume} {26}},\ \bibinfo
  {pages} {26776} (\bibinfo {year} {2018})}\BibitemShut {NoStop}%
\bibitem [{\citenamefont {Yoon}\ \emph {et~al.}(2019)\citenamefont {Yoon},
  \citenamefont {Jeon}, \citenamefont {Shin}, \citenamefont {Lee},
  \citenamefont {Lee}, \citenamefont {Choi}, \citenamefont {Kim}, \citenamefont
  {Sung},\ and\ \citenamefont {Nam}}]{Yoon:OE:2019}%
  \BibitemOpen
  \bibfield  {author} {\bibinfo {author} {\bibfnamefont {J.~W.}\ \bibnamefont
  {Yoon}}, \bibinfo {author} {\bibfnamefont {C.}~\bibnamefont {Jeon}}, \bibinfo
  {author} {\bibfnamefont {J.}~\bibnamefont {Shin}}, \bibinfo {author}
  {\bibfnamefont {S.~K.}\ \bibnamefont {Lee}}, \bibinfo {author} {\bibfnamefont
  {H.~W.}\ \bibnamefont {Lee}}, \bibinfo {author} {\bibfnamefont {I.~W.}\
  \bibnamefont {Choi}}, \bibinfo {author} {\bibfnamefont {H.~T.}\ \bibnamefont
  {Kim}}, \bibinfo {author} {\bibfnamefont {J.~H.}\ \bibnamefont {Sung}},\ and\
  \bibinfo {author} {\bibfnamefont {C.~H.}\ \bibnamefont {Nam}},\ }\bibfield
  {title} {\bibinfo {title} {Achieving the laser intensity of
  $5.5\times10^{22}~\mathrm{W/cm^2}$ with a wavefront-corrected {multi-PW}
  laser},\ }\href {https://doi.org/10.1364/OE.27.020412} {\bibfield  {journal}
  {\bibinfo  {journal} {Opt. Express}\ }\textbf {\bibinfo {volume} {27}},\
  \bibinfo {pages} {20412} (\bibinfo {year} {2019})}\BibitemShut {NoStop}%
\bibitem [{\citenamefont {Wagner}\ \emph {et~al.}(2016)\citenamefont {Wagner},
  \citenamefont {Deppert}, \citenamefont {Brabetz}, \citenamefont {Fiala},
  \citenamefont {Kleinschmidt}, \citenamefont {Poth}, \citenamefont {Schanz},
  \citenamefont {Tebartz}, \citenamefont {Zielbauer}, \citenamefont {Roth}
  \emph {et~al.}}]{Wagner:2016:PhysRevLett.116.205002}%
  \BibitemOpen
  \bibfield  {author} {\bibinfo {author} {\bibfnamefont {F.}~\bibnamefont
  {Wagner}}, \bibinfo {author} {\bibfnamefont {O.}~\bibnamefont {Deppert}},
  \bibinfo {author} {\bibfnamefont {C.}~\bibnamefont {Brabetz}}, \bibinfo
  {author} {\bibfnamefont {P.}~\bibnamefont {Fiala}}, \bibinfo {author}
  {\bibfnamefont {A.}~\bibnamefont {Kleinschmidt}}, \bibinfo {author}
  {\bibfnamefont {P.}~\bibnamefont {Poth}}, \bibinfo {author} {\bibfnamefont
  {V.~A.}\ \bibnamefont {Schanz}}, \bibinfo {author} {\bibfnamefont
  {A.}~\bibnamefont {Tebartz}}, \bibinfo {author} {\bibfnamefont
  {B.}~\bibnamefont {Zielbauer}}, \bibinfo {author} {\bibfnamefont
  {M.}~\bibnamefont {Roth}}, \emph {et~al.},\ }\bibfield  {title} {\bibinfo
  {title} {Maximum proton energy above 85 mev from the relativistic interaction
  of laser pulses with micrometer thick ${\mathrm{ch}}_{2}$ targets},\ }\href
  {https://doi.org/10.1103/PhysRevLett.116.205002} {\bibfield  {journal}
  {\bibinfo  {journal} {Phys. Rev. Lett.}\ }\textbf {\bibinfo {volume} {116}},\
  \bibinfo {pages} {205002} (\bibinfo {year} {2016})}\BibitemShut {NoStop}%
\bibitem [{\citenamefont {Kim}\ \emph {et~al.}(2016)\citenamefont {Kim},
  \citenamefont {Pae}, \citenamefont {Choi}, \citenamefont {Lee}, \citenamefont
  {Kim}, \citenamefont {Singhal}, \citenamefont {Sung}, \citenamefont {Lee},
  \citenamefont {Lee}, \citenamefont {Nickles} \emph {et~al.}}]{Kim::POP:2016}%
  \BibitemOpen
  \bibfield  {author} {\bibinfo {author} {\bibfnamefont {I.~J.}\ \bibnamefont
  {Kim}}, \bibinfo {author} {\bibfnamefont {K.~H.}\ \bibnamefont {Pae}},
  \bibinfo {author} {\bibfnamefont {I.~W.}\ \bibnamefont {Choi}}, \bibinfo
  {author} {\bibfnamefont {C.-L.}\ \bibnamefont {Lee}}, \bibinfo {author}
  {\bibfnamefont {H.~T.}\ \bibnamefont {Kim}}, \bibinfo {author} {\bibfnamefont
  {H.}~\bibnamefont {Singhal}}, \bibinfo {author} {\bibfnamefont {J.~H.}\
  \bibnamefont {Sung}}, \bibinfo {author} {\bibfnamefont {S.~K.}\ \bibnamefont
  {Lee}}, \bibinfo {author} {\bibfnamefont {H.~W.}\ \bibnamefont {Lee}},
  \bibinfo {author} {\bibfnamefont {P.~V.}\ \bibnamefont {Nickles}}, \emph
  {et~al.},\ }\bibfield  {title} {\bibinfo {title} {Radiation pressure
  acceleration of protons to 93 {MeV} with circularly polarized petawatt laser
  pulses},\ }\href {https://doi.org/10.1063/1.4958654} {\bibfield  {journal}
  {\bibinfo  {journal} {Phys. Plasmas}\ }\textbf {\bibinfo {volume} {23}},\
  \bibinfo {pages} {070701} (\bibinfo {year} {2016})}\BibitemShut {NoStop}%
\bibitem [{\citenamefont {Zhang}\ \emph {et~al.}(2017)\citenamefont {Zhang},
  \citenamefont {Shen}, \citenamefont {Wang}, \citenamefont {Zhai},
  \citenamefont {Li}, \citenamefont {Lu}, \citenamefont {Li}, \citenamefont
  {Xu}, \citenamefont {Wang}, \citenamefont {Liang} \emph
  {et~al.}}]{Zhang:2017:PhysRevLett.119.164801}%
  \BibitemOpen
  \bibfield  {author} {\bibinfo {author} {\bibfnamefont {H.}~\bibnamefont
  {Zhang}}, \bibinfo {author} {\bibfnamefont {B.~F.}\ \bibnamefont {Shen}},
  \bibinfo {author} {\bibfnamefont {W.~P.}\ \bibnamefont {Wang}}, \bibinfo
  {author} {\bibfnamefont {S.~H.}\ \bibnamefont {Zhai}}, \bibinfo {author}
  {\bibfnamefont {S.~S.}\ \bibnamefont {Li}}, \bibinfo {author} {\bibfnamefont
  {X.~M.}\ \bibnamefont {Lu}}, \bibinfo {author} {\bibfnamefont {J.~F.}\
  \bibnamefont {Li}}, \bibinfo {author} {\bibfnamefont {R.~J.}\ \bibnamefont
  {Xu}}, \bibinfo {author} {\bibfnamefont {X.~L.}\ \bibnamefont {Wang}},
  \bibinfo {author} {\bibfnamefont {X.~Y.}\ \bibnamefont {Liang}}, \emph
  {et~al.},\ }\bibfield  {title} {\bibinfo {title} {Collisionless shock
  acceleration of high-flux quasimonoenergetic proton beams driven by
  circularly polarized laser pulses},\ }\href
  {https://doi.org/10.1103/PhysRevLett.119.164801} {\bibfield  {journal}
  {\bibinfo  {journal} {Phys. Rev. Lett.}\ }\textbf {\bibinfo {volume} {119}},\
  \bibinfo {pages} {164801} (\bibinfo {year} {2017})}\BibitemShut {NoStop}%
\bibitem [{\citenamefont {Hilz}\ \emph {et~al.}(2018)\citenamefont {Hilz},
  \citenamefont {Ostermayr}, \citenamefont {Huebl}, \citenamefont {Bagnoud},
  \citenamefont {Borm}, \citenamefont {Bussmann}, \citenamefont {Gallei},
  \citenamefont {Gebhard}, \citenamefont {Haffa}, \citenamefont {Hartmann}
  \emph {et~al.}}]{Hilz:NatC:2018}%
  \BibitemOpen
  \bibfield  {author} {\bibinfo {author} {\bibfnamefont {P.}~\bibnamefont
  {Hilz}}, \bibinfo {author} {\bibfnamefont {T.~M.}\ \bibnamefont {Ostermayr}},
  \bibinfo {author} {\bibfnamefont {A.}~\bibnamefont {Huebl}}, \bibinfo
  {author} {\bibfnamefont {V.}~\bibnamefont {Bagnoud}}, \bibinfo {author}
  {\bibfnamefont {B.}~\bibnamefont {Borm}}, \bibinfo {author} {\bibfnamefont
  {M.}~\bibnamefont {Bussmann}}, \bibinfo {author} {\bibfnamefont
  {M.}~\bibnamefont {Gallei}}, \bibinfo {author} {\bibfnamefont
  {J.}~\bibnamefont {Gebhard}}, \bibinfo {author} {\bibfnamefont
  {D.}~\bibnamefont {Haffa}}, \bibinfo {author} {\bibfnamefont
  {J.}~\bibnamefont {Hartmann}}, \emph {et~al.},\ }\bibfield  {title} {\bibinfo
  {title} {Isolated proton bunch acceleration by a petawatt laser pulse},\
  }\href {https://doi.org/10.1038/s41467-017-02663-1} {\bibfield  {journal}
  {\bibinfo  {journal} {Nat. Commun.}\ }\textbf {\bibinfo {volume} {9}},\
  \bibinfo {pages} {423} (\bibinfo {year} {2018})}\BibitemShut {NoStop}%
\bibitem [{\citenamefont {Higginson}\ \emph {et~al.}(2018)\citenamefont
  {Higginson}, \citenamefont {Gray}, \citenamefont {King}, \citenamefont
  {Dance}, \citenamefont {Williamson}, \citenamefont {Butler}, \citenamefont
  {Wilson}, \citenamefont {Capdessus}, \citenamefont {Armstrong}, \citenamefont
  {Green} \emph {et~al.}}]{Higginson:NC:2018}%
  \BibitemOpen
  \bibfield  {author} {\bibinfo {author} {\bibfnamefont {A.}~\bibnamefont
  {Higginson}}, \bibinfo {author} {\bibfnamefont {R.~J.}\ \bibnamefont {Gray}},
  \bibinfo {author} {\bibfnamefont {M.}~\bibnamefont {King}}, \bibinfo {author}
  {\bibfnamefont {R.~J.}\ \bibnamefont {Dance}}, \bibinfo {author}
  {\bibfnamefont {S.~D.~R.}\ \bibnamefont {Williamson}}, \bibinfo {author}
  {\bibfnamefont {N.~M.~H.}\ \bibnamefont {Butler}}, \bibinfo {author}
  {\bibfnamefont {R.}~\bibnamefont {Wilson}}, \bibinfo {author} {\bibfnamefont
  {R.}~\bibnamefont {Capdessus}}, \bibinfo {author} {\bibfnamefont
  {C.}~\bibnamefont {Armstrong}}, \bibinfo {author} {\bibfnamefont {J.~S.}\
  \bibnamefont {Green}}, \emph {et~al.},\ }\bibfield  {title} {\bibinfo {title}
  {Near-100 {MeV} protons via a laser-driven transparency-enhanced hybrid
  acceleration scheme},\ }\href {https://doi.org/10.1038/s41467-018-03063-9}
  {\bibfield  {journal} {\bibinfo  {journal} {Nature Communications}\ }\textbf
  {\bibinfo {volume} {9}},\ \bibinfo {pages} {724} (\bibinfo {year}
  {2018})}\BibitemShut {NoStop}%
\bibitem [{\citenamefont {Ma}\ \emph {et~al.}(2019)\citenamefont {Ma},
  \citenamefont {Kim}, \citenamefont {Yu}, \citenamefont {Choi}, \citenamefont
  {Singh}, \citenamefont {Lee}, \citenamefont {Sung}, \citenamefont {Lee},
  \citenamefont {Lin}, \citenamefont {Liao} \emph
  {et~al.}}]{Ma:2019:PhysRevLett.122.014803}%
  \BibitemOpen
  \bibfield  {author} {\bibinfo {author} {\bibfnamefont {W.~J.}\ \bibnamefont
  {Ma}}, \bibinfo {author} {\bibfnamefont {I.~J.}\ \bibnamefont {Kim}},
  \bibinfo {author} {\bibfnamefont {J.~Q.}\ \bibnamefont {Yu}}, \bibinfo
  {author} {\bibfnamefont {I.~W.}\ \bibnamefont {Choi}}, \bibinfo {author}
  {\bibfnamefont {P.~K.}\ \bibnamefont {Singh}}, \bibinfo {author}
  {\bibfnamefont {H.~W.}\ \bibnamefont {Lee}}, \bibinfo {author} {\bibfnamefont
  {J.~H.}\ \bibnamefont {Sung}}, \bibinfo {author} {\bibfnamefont {S.~K.}\
  \bibnamefont {Lee}}, \bibinfo {author} {\bibfnamefont {C.}~\bibnamefont
  {Lin}}, \bibinfo {author} {\bibfnamefont {Q.}~\bibnamefont {Liao}}, \emph
  {et~al.},\ }\bibfield  {title} {\bibinfo {title} {Laser acceleration of
  highly energetic carbon ions using a double-layer target composed of slightly
  underdense plasma and ultrathin foil},\ }\href
  {https://doi.org/10.1103/PhysRevLett.122.014803} {\bibfield  {journal}
  {\bibinfo  {journal} {Phys. Rev. Lett.}\ }\textbf {\bibinfo {volume} {122}},\
  \bibinfo {pages} {014803} (\bibinfo {year} {2019})}\BibitemShut {NoStop}%
\bibitem [{\citenamefont {Kraft}\ \emph {et~al.}(2018)\citenamefont {Kraft},
  \citenamefont {Obst}, \citenamefont {Metzkes-Ng}, \citenamefont
  {Schlenvoigt}, \citenamefont {Zeil}, \citenamefont {Michaux}, \citenamefont
  {Chatain}, \citenamefont {Perin}, \citenamefont {Chen}, \citenamefont
  {Fuchs}, \citenamefont {Gauthier}, \citenamefont {Cowan},\ and\ \citenamefont
  {Schramm}}]{Kraft:PPCF:2018}%
  \BibitemOpen
  \bibfield  {author} {\bibinfo {author} {\bibfnamefont {S.~D.}\ \bibnamefont
  {Kraft}}, \bibinfo {author} {\bibfnamefont {L.}~\bibnamefont {Obst}},
  \bibinfo {author} {\bibfnamefont {J.}~\bibnamefont {Metzkes-Ng}}, \bibinfo
  {author} {\bibfnamefont {H.-P.}\ \bibnamefont {Schlenvoigt}}, \bibinfo
  {author} {\bibfnamefont {K.}~\bibnamefont {Zeil}}, \bibinfo {author}
  {\bibfnamefont {S.}~\bibnamefont {Michaux}}, \bibinfo {author} {\bibfnamefont
  {D.}~\bibnamefont {Chatain}}, \bibinfo {author} {\bibfnamefont {J.-P.}\
  \bibnamefont {Perin}}, \bibinfo {author} {\bibfnamefont {S.~N.}\ \bibnamefont
  {Chen}}, \bibinfo {author} {\bibfnamefont {J.}~\bibnamefont {Fuchs}},
  \bibinfo {author} {\bibfnamefont {M.}~\bibnamefont {Gauthier}}, \bibinfo
  {author} {\bibfnamefont {T.~E.}\ \bibnamefont {Cowan}},\ and\ \bibinfo
  {author} {\bibfnamefont {U.}~\bibnamefont {Schramm}},\ }\bibfield  {title}
  {\bibinfo {title} {First demonstration of multi-{MeV} proton acceleration
  from a cryogenic hydrogen ribbon target},\ }\href
  {https://doi.org/10.1088/1361-6587/aaae38} {\bibfield  {journal} {\bibinfo
  {journal} {Plasma Phys. Control. Fusion}\ }\textbf {\bibinfo {volume} {60}},\
  \bibinfo {pages} {044010} (\bibinfo {year} {2018})}\BibitemShut {NoStop}%
\bibitem [{\citenamefont {H{\" u}tzen}\ \emph {et~al.}(2019)\citenamefont {H{\"
  u}tzen}, \citenamefont {Thomas}, \citenamefont {B{\" o}ker}, \citenamefont
  {Engels}, \citenamefont {Gebel}, \citenamefont {Lehrach}, \citenamefont
  {Pukhov}, \citenamefont {Rakitzis}, \citenamefont {Sofikitis}, \citenamefont
  {B{\" u}scher},\ and\ \citenamefont {et~al.}}]{huetzenHPLSE19}%
  \BibitemOpen
  \bibfield  {author} {\bibinfo {author} {\bibfnamefont {A.}~\bibnamefont {H{\"
  u}tzen}}, \bibinfo {author} {\bibfnamefont {J.}~\bibnamefont {Thomas}},
  \bibinfo {author} {\bibfnamefont {J.}~\bibnamefont {B{\" o}ker}}, \bibinfo
  {author} {\bibfnamefont {R.}~\bibnamefont {Engels}}, \bibinfo {author}
  {\bibfnamefont {R.}~\bibnamefont {Gebel}}, \bibinfo {author} {\bibfnamefont
  {A.}~\bibnamefont {Lehrach}}, \bibinfo {author} {\bibfnamefont
  {A.}~\bibnamefont {Pukhov}}, \bibinfo {author} {\bibfnamefont {T.~P.}\
  \bibnamefont {Rakitzis}}, \bibinfo {author} {\bibfnamefont {D.}~\bibnamefont
  {Sofikitis}}, \bibinfo {author} {\bibfnamefont {M.}~\bibnamefont {B{\"
  u}scher}},\ and\ \bibinfo {author} {\bibnamefont {et~al.}},\ }\bibfield
  {title} {\bibinfo {title} {Polarized proton beams from laser-induced
  plasmas},\ }\href {https://doi.org/10.1017/hpl.2018.73} {\bibfield  {journal}
  {\bibinfo  {journal} {High Power Laser Sci. Eng.}\ }\textbf {\bibinfo
  {volume} {7}},\ \bibinfo {pages} {e16} (\bibinfo {year} {2019})}\BibitemShut
  {NoStop}%
\bibitem [{\citenamefont {B\"uscher}\ \emph {et~al.}(2019)\citenamefont
  {B\"uscher}, \citenamefont {H\"utzen}, \citenamefont {Engin}, \citenamefont
  {Thomas}, \citenamefont {Pukhov}, \citenamefont {B\"oker}, \citenamefont
  {Gebel}, \citenamefont {Lehrach}, \citenamefont {Engels}, \citenamefont
  {Peter~Rakitzis},\ and\ \citenamefont {Sofikitis}}]{Buscher:IJMPA:2019}%
  \BibitemOpen
  \bibfield  {author} {\bibinfo {author} {\bibfnamefont {M.}~\bibnamefont
  {B\"uscher}}, \bibinfo {author} {\bibfnamefont {A.}~\bibnamefont {H\"utzen}},
  \bibinfo {author} {\bibfnamefont {I.}~\bibnamefont {Engin}}, \bibinfo
  {author} {\bibfnamefont {J.}~\bibnamefont {Thomas}}, \bibinfo {author}
  {\bibfnamefont {A.}~\bibnamefont {Pukhov}}, \bibinfo {author} {\bibfnamefont
  {J.}~\bibnamefont {B\"oker}}, \bibinfo {author} {\bibfnamefont
  {R.}~\bibnamefont {Gebel}}, \bibinfo {author} {\bibfnamefont
  {A.}~\bibnamefont {Lehrach}}, \bibinfo {author} {\bibfnamefont
  {R.}~\bibnamefont {Engels}}, \bibinfo {author} {\bibfnamefont
  {T.}~\bibnamefont {Peter~Rakitzis}},\ and\ \bibinfo {author} {\bibfnamefont
  {D.}~\bibnamefont {Sofikitis}},\ }\bibfield  {title} {\bibinfo {title}
  {Polarized proton beams from a laser-plasma accelerator},\ }\href
  {https://doi.org/10.1142/S0217751X19420284} {\bibfield  {journal} {\bibinfo
  {journal} {Int. J. Mod. Phys. A}\ ,\ \bibinfo {pages} {1942028}} (\bibinfo
  {year} {2019})}\BibitemShut {NoStop}%
\bibitem [{\citenamefont {Li}\ \emph {et~al.}(2019)\citenamefont {Li},
  \citenamefont {Shaisultanov}, \citenamefont {Hatsagortsyan}, \citenamefont
  {Wan}, \citenamefont {Keitel},\ and\ \citenamefont
  {Li}}]{Li:2019:PhysRevLett.122.154801}%
  \BibitemOpen
  \bibfield  {author} {\bibinfo {author} {\bibfnamefont {Y.-F.}\ \bibnamefont
  {Li}}, \bibinfo {author} {\bibfnamefont {R.}~\bibnamefont {Shaisultanov}},
  \bibinfo {author} {\bibfnamefont {K.~Z.}\ \bibnamefont {Hatsagortsyan}},
  \bibinfo {author} {\bibfnamefont {F.}~\bibnamefont {Wan}}, \bibinfo {author}
  {\bibfnamefont {C.~H.}\ \bibnamefont {Keitel}},\ and\ \bibinfo {author}
  {\bibfnamefont {J.-X.}\ \bibnamefont {Li}},\ }\bibfield  {title} {\bibinfo
  {title} {Ultrarelativistic electron-beam polarization in single-shot
  interaction with an ultraintense laser pulse},\ }\href
  {https://doi.org/10.1103/PhysRevLett.122.154801} {\bibfield  {journal}
  {\bibinfo  {journal} {Phys. Rev. Lett.}\ }\textbf {\bibinfo {volume} {122}},\
  \bibinfo {pages} {154801} (\bibinfo {year} {2019})}\BibitemShut {NoStop}%
\bibitem [{\citenamefont {Wen}\ \emph {et~al.}(2019)\citenamefont {Wen},
  \citenamefont {Tamburini},\ and\ \citenamefont
  {Keitel}}]{Wen:2019:PhysRevLett.122.214801}%
  \BibitemOpen
  \bibfield  {author} {\bibinfo {author} {\bibfnamefont {M.}~\bibnamefont
  {Wen}}, \bibinfo {author} {\bibfnamefont {M.}~\bibnamefont {Tamburini}},\
  and\ \bibinfo {author} {\bibfnamefont {C.~H.}\ \bibnamefont {Keitel}},\
  }\bibfield  {title} {\bibinfo {title} {Polarized laser-wakefield-accelerated
  kiloampere electron beams},\ }\href
  {https://doi.org/10.1103/PhysRevLett.122.214801} {\bibfield  {journal}
  {\bibinfo  {journal} {Phys. Rev. Lett.}\ }\textbf {\bibinfo {volume} {122}},\
  \bibinfo {pages} {214801} (\bibinfo {year} {2019})}\BibitemShut {NoStop}%
\bibitem [{\citenamefont {Wu}\ \emph {et~al.}(2019)\citenamefont {Wu},
  \citenamefont {Ji}, \citenamefont {Geng}, \citenamefont {Yu}, \citenamefont
  {Wang}, \citenamefont {Feng}, \citenamefont {Guo}, \citenamefont {Wang},
  \citenamefont {Qin}, \citenamefont {Yan}, \citenamefont {Zhang},
  \citenamefont {Thomas}, \citenamefont {H\"utzen}, \citenamefont {Pukhov},
  \citenamefont {B\"uscher}, \citenamefont {Shen},\ and\ \citenamefont
  {Li}}]{Wu:2019:PhysRevE.100.043202}%
  \BibitemOpen
  \bibfield  {author} {\bibinfo {author} {\bibfnamefont {Y.}~\bibnamefont
  {Wu}}, \bibinfo {author} {\bibfnamefont {L.}~\bibnamefont {Ji}}, \bibinfo
  {author} {\bibfnamefont {X.}~\bibnamefont {Geng}}, \bibinfo {author}
  {\bibfnamefont {Q.}~\bibnamefont {Yu}}, \bibinfo {author} {\bibfnamefont
  {N.}~\bibnamefont {Wang}}, \bibinfo {author} {\bibfnamefont {B.}~\bibnamefont
  {Feng}}, \bibinfo {author} {\bibfnamefont {Z.}~\bibnamefont {Guo}}, \bibinfo
  {author} {\bibfnamefont {W.}~\bibnamefont {Wang}}, \bibinfo {author}
  {\bibfnamefont {C.}~\bibnamefont {Qin}}, \bibinfo {author} {\bibfnamefont
  {X.}~\bibnamefont {Yan}}, \bibinfo {author} {\bibfnamefont {L.}~\bibnamefont
  {Zhang}}, \bibinfo {author} {\bibfnamefont {J.}~\bibnamefont {Thomas}},
  \bibinfo {author} {\bibfnamefont {A.}~\bibnamefont {H\"utzen}}, \bibinfo
  {author} {\bibfnamefont {A.}~\bibnamefont {Pukhov}}, \bibinfo {author}
  {\bibfnamefont {M.}~\bibnamefont {B\"uscher}}, \bibinfo {author}
  {\bibfnamefont {B.}~\bibnamefont {Shen}},\ and\ \bibinfo {author}
  {\bibfnamefont {R.}~\bibnamefont {Li}},\ }\bibfield  {title} {\bibinfo
  {title} {Polarized electron acceleration in beam-driven plasma wakefield
  based on density down-ramp injection},\ }\href
  {https://doi.org/10.1103/PhysRevE.100.043202} {\bibfield  {journal} {\bibinfo
   {journal} {Phys. Rev. E}\ }\textbf {\bibinfo {volume} {100}},\ \bibinfo
  {pages} {043202} (\bibinfo {year} {2019})}\BibitemShut {NoStop}%
\bibitem [{\citenamefont {Thomas}\ \emph {et~al.}(2020)\citenamefont {Thomas},
  \citenamefont {H\"utzen}, \citenamefont {Lehrach}, \citenamefont {Pukhov},
  \citenamefont {Ji}, \citenamefont {Wu}, \citenamefont {Geng},\ and\
  \citenamefont {B\"uscher}}]{Thomas:2020}%
  \BibitemOpen
  \bibfield  {author} {\bibinfo {author} {\bibfnamefont {J.}~\bibnamefont
  {Thomas}}, \bibinfo {author} {\bibfnamefont {A.}~\bibnamefont {H\"utzen}},
  \bibinfo {author} {\bibfnamefont {A.}~\bibnamefont {Lehrach}}, \bibinfo
  {author} {\bibfnamefont {A.}~\bibnamefont {Pukhov}}, \bibinfo {author}
  {\bibfnamefont {L.}~\bibnamefont {Ji}}, \bibinfo {author} {\bibfnamefont
  {Y.}~\bibnamefont {Wu}}, \bibinfo {author} {\bibfnamefont {X.}~\bibnamefont
  {Geng}},\ and\ \bibinfo {author} {\bibfnamefont {M.}~\bibnamefont
  {B\"uscher}},\ }\bibfield  {title} {\bibinfo {title} {Scaling laws for the
  (de-)polarization time of relativistic particle beams in strong fields},\
  }\href {https://arxiv.org/abs/2001.07084} {\bibfield  {journal} {\bibinfo
  {journal} {arXiv:2001.07084}\ } (\bibinfo {year} {2020})}\BibitemShut
  {NoStop}%
\bibitem [{\citenamefont {Sofikitis}\ \emph {et~al.}(2018)\citenamefont
  {Sofikitis}, \citenamefont {Kannis}, \citenamefont {Boulogiannis},\ and\
  \citenamefont {Rakitzis}}]{Sofikitis:2018:PhysRevLett.121.083001}%
  \BibitemOpen
  \bibfield  {author} {\bibinfo {author} {\bibfnamefont {D.}~\bibnamefont
  {Sofikitis}}, \bibinfo {author} {\bibfnamefont {C.~S.}\ \bibnamefont
  {Kannis}}, \bibinfo {author} {\bibfnamefont {G.~K.}\ \bibnamefont
  {Boulogiannis}},\ and\ \bibinfo {author} {\bibfnamefont {T.~P.}\ \bibnamefont
  {Rakitzis}},\ }\bibfield  {title} {\bibinfo {title} {{Ultrahigh-Density
  Spin-Polarized H and D Observed via Magnetization Quantum Beats}},\ }\href
  {https://doi.org/10.1103/PhysRevLett.121.083001} {\bibfield  {journal}
  {\bibinfo  {journal} {Phys. Rev. Lett.}\ }\textbf {\bibinfo {volume} {121}},\
  \bibinfo {pages} {083001} (\bibinfo {year} {2018})}\BibitemShut {NoStop}%
\bibitem [{\citenamefont {Shen}\ \emph {et~al.}(2007)\citenamefont {Shen},
  \citenamefont {Li}, \citenamefont {Yu},\ and\ \citenamefont
  {Cary}}]{Shen:2007:PhysRevE.76.055402}%
  \BibitemOpen
  \bibfield  {author} {\bibinfo {author} {\bibfnamefont {B.}~\bibnamefont
  {Shen}}, \bibinfo {author} {\bibfnamefont {Y.}~\bibnamefont {Li}}, \bibinfo
  {author} {\bibfnamefont {M.~Y.}\ \bibnamefont {Yu}},\ and\ \bibinfo {author}
  {\bibfnamefont {J.}~\bibnamefont {Cary}},\ }\bibfield  {title} {\bibinfo
  {title} {Bubble regime for ion acceleration in a laser-driven plasma},\
  }\href {https://doi.org/10.1103/PhysRevE.76.055402} {\bibfield  {journal}
  {\bibinfo  {journal} {Phys. Rev. E}\ }\textbf {\bibinfo {volume} {76}},\
  \bibinfo {pages} {055402} (\bibinfo {year} {2007})}\BibitemShut {NoStop}%
\bibitem [{\citenamefont {Zhang}\ \emph {et~al.}(2014)\citenamefont {Zhang},
  \citenamefont {Shen}, \citenamefont {Zhang}, \citenamefont {Xu},
  \citenamefont {Wang}, \citenamefont {Wang}, \citenamefont {Yi},\ and\
  \citenamefont {Shi}}]{Zhang:NJP:2014}%
  \BibitemOpen
  \bibfield  {author} {\bibinfo {author} {\bibfnamefont {X.}~\bibnamefont
  {Zhang}}, \bibinfo {author} {\bibfnamefont {B.}~\bibnamefont {Shen}},
  \bibinfo {author} {\bibfnamefont {L.}~\bibnamefont {Zhang}}, \bibinfo
  {author} {\bibfnamefont {J.}~\bibnamefont {Xu}}, \bibinfo {author}
  {\bibfnamefont {X.}~\bibnamefont {Wang}}, \bibinfo {author} {\bibfnamefont
  {W.}~\bibnamefont {Wang}}, \bibinfo {author} {\bibfnamefont {L.}~\bibnamefont
  {Yi}},\ and\ \bibinfo {author} {\bibfnamefont {Y.}~\bibnamefont {Shi}},\
  }\bibfield  {title} {\bibinfo {title} {Proton acceleration in underdense
  plasma by ultraintense laguerre{\textendash}gaussian laser pulse},\ }\href
  {https://doi.org/10.1088/1367-2630/16/12/123051} {\bibfield  {journal}
  {\bibinfo  {journal} {New J. Phys.}\ }\textbf {\bibinfo {volume} {16}},\
  \bibinfo {pages} {123051} (\bibinfo {year} {2014})}\BibitemShut {NoStop}%
\bibitem [{\citenamefont {Wan}\ \emph {et~al.}(2019)\citenamefont {Wan},
  \citenamefont {Andriyash}, \citenamefont {Hua}, \citenamefont {Pai},
  \citenamefont {Lu}, \citenamefont {Mori}, \citenamefont {Joshi},\ and\
  \citenamefont {Malka}}]{Wan:2019:PhysRevAccelBeams.22.021301}%
  \BibitemOpen
  \bibfield  {author} {\bibinfo {author} {\bibfnamefont {Y.}~\bibnamefont
  {Wan}}, \bibinfo {author} {\bibfnamefont {I.~A.}\ \bibnamefont {Andriyash}},
  \bibinfo {author} {\bibfnamefont {J.~F.}\ \bibnamefont {Hua}}, \bibinfo
  {author} {\bibfnamefont {C.-H.}\ \bibnamefont {Pai}}, \bibinfo {author}
  {\bibfnamefont {W.}~\bibnamefont {Lu}}, \bibinfo {author} {\bibfnamefont
  {W.~B.}\ \bibnamefont {Mori}}, \bibinfo {author} {\bibfnamefont
  {C.}~\bibnamefont {Joshi}},\ and\ \bibinfo {author} {\bibfnamefont
  {V.}~\bibnamefont {Malka}},\ }\bibfield  {title} {\bibinfo {title} {Two-stage
  laser acceleration of high quality protons using a tailored density plasma},\
  }\href {https://doi.org/10.1103/PhysRevAccelBeams.22.021301} {\bibfield
  {journal} {\bibinfo  {journal} {Phys. Rev. Accel. Beams}\ }\textbf {\bibinfo
  {volume} {22}},\ \bibinfo {pages} {021301} (\bibinfo {year}
  {2019})}\BibitemShut {NoStop}%
\bibitem [{\citenamefont {Engin}\ \emph {et~al.}(2019)\citenamefont {Engin},
  \citenamefont {Chitgar}, \citenamefont {Deppert}, \citenamefont {Lucchio},
  \citenamefont {Engels}, \citenamefont {Fedorets}, \citenamefont {Frydrych},
  \citenamefont {Gibbon}, \citenamefont {Kleinschmidt}, \citenamefont
  {Lehrach}, \citenamefont {Maier}, \citenamefont {Prasuhn}, \citenamefont
  {Roth}, \citenamefont {Schlüter}, \citenamefont {Schneider}, \citenamefont
  {Stöhlker}, \citenamefont {Strathmann},\ and\ \citenamefont
  {Büscher}}]{Engin:PPCF:2019}%
  \BibitemOpen
  \bibfield  {author} {\bibinfo {author} {\bibfnamefont {I.}~\bibnamefont
  {Engin}}, \bibinfo {author} {\bibfnamefont {Z.~M.}\ \bibnamefont {Chitgar}},
  \bibinfo {author} {\bibfnamefont {O.}~\bibnamefont {Deppert}}, \bibinfo
  {author} {\bibfnamefont {L.~D.}\ \bibnamefont {Lucchio}}, \bibinfo {author}
  {\bibfnamefont {R.}~\bibnamefont {Engels}}, \bibinfo {author} {\bibfnamefont
  {P.}~\bibnamefont {Fedorets}}, \bibinfo {author} {\bibfnamefont
  {S.}~\bibnamefont {Frydrych}}, \bibinfo {author} {\bibfnamefont
  {P.}~\bibnamefont {Gibbon}}, \bibinfo {author} {\bibfnamefont
  {A.}~\bibnamefont {Kleinschmidt}}, \bibinfo {author} {\bibfnamefont
  {A.}~\bibnamefont {Lehrach}}, \bibinfo {author} {\bibfnamefont
  {R.}~\bibnamefont {Maier}}, \bibinfo {author} {\bibfnamefont
  {D.}~\bibnamefont {Prasuhn}}, \bibinfo {author} {\bibfnamefont
  {M.}~\bibnamefont {Roth}}, \bibinfo {author} {\bibfnamefont {F.}~\bibnamefont
  {Schlüter}}, \bibinfo {author} {\bibfnamefont {C.~M.}\ \bibnamefont
  {Schneider}}, \bibinfo {author} {\bibfnamefont {T.}~\bibnamefont
  {Stöhlker}}, \bibinfo {author} {\bibfnamefont {K.}~\bibnamefont
  {Strathmann}},\ and\ \bibinfo {author} {\bibfnamefont {M.}~\bibnamefont
  {Büscher}},\ }\bibfield  {title} {\bibinfo {title} {Laser-induced
  acceleration of helium ions from unpolarized gas jets},\ }\href
  {https://doi.org/10.1088/1361-6587/ab4613} {\bibfield  {journal} {\bibinfo
  {journal} {Plasma Phys. Control. Fusion}\ }\textbf {\bibinfo {volume} {61}},\
  \bibinfo {pages} {115012} (\bibinfo {year} {2019})}\BibitemShut {NoStop}%
\bibitem [{\citenamefont {Arber}\ \emph {et~al.}(2015)\citenamefont {Arber},
  \citenamefont {Bennett}, \citenamefont {Brady}, \citenamefont
  {Lawrence-Douglas}, \citenamefont {Ramsay}, \citenamefont {Sircombe},
  \citenamefont {Gillies}, \citenamefont {Evans}, \citenamefont {Schmitz},
  \citenamefont {Bell} \emph {et~al.}}]{Arber:epoch:2015hc}%
  \BibitemOpen
  \bibfield  {author} {\bibinfo {author} {\bibfnamefont {T.~D.}\ \bibnamefont
  {Arber}}, \bibinfo {author} {\bibfnamefont {K.}~\bibnamefont {Bennett}},
  \bibinfo {author} {\bibfnamefont {C.~S.}\ \bibnamefont {Brady}}, \bibinfo
  {author} {\bibfnamefont {A.}~\bibnamefont {Lawrence-Douglas}}, \bibinfo
  {author} {\bibfnamefont {M.~G.}\ \bibnamefont {Ramsay}}, \bibinfo {author}
  {\bibfnamefont {N.~J.}\ \bibnamefont {Sircombe}}, \bibinfo {author}
  {\bibfnamefont {P.}~\bibnamefont {Gillies}}, \bibinfo {author} {\bibfnamefont
  {R.~G.}\ \bibnamefont {Evans}}, \bibinfo {author} {\bibfnamefont
  {H.}~\bibnamefont {Schmitz}}, \bibinfo {author} {\bibfnamefont {A.~R.}\
  \bibnamefont {Bell}}, \emph {et~al.},\ }\bibfield  {title} {\bibinfo {title}
  {Contemporary particle-in-cell approach to laser-plasma modelling},\ }\href
  {http://stacks.iop.org/0741-3335/57/i=11/a=113001} {\bibfield  {journal}
  {\bibinfo  {journal} {Plasma Phys. Control. Fusion}\ }\textbf {\bibinfo
  {volume} {57}},\ \bibinfo {pages} {113001} (\bibinfo {year}
  {2015})}\BibitemShut {NoStop}%
\bibitem [{\citenamefont {Zamanian}\ \emph {et~al.}(2010)\citenamefont
  {Zamanian}, \citenamefont {Stefan}, \citenamefont {Marklund},\ and\
  \citenamefont {Brodin}}]{Zamanian:2010:POP}%
  \BibitemOpen
  \bibfield  {author} {\bibinfo {author} {\bibfnamefont {J.}~\bibnamefont
  {Zamanian}}, \bibinfo {author} {\bibfnamefont {M.}~\bibnamefont {Stefan}},
  \bibinfo {author} {\bibfnamefont {M.}~\bibnamefont {Marklund}},\ and\
  \bibinfo {author} {\bibfnamefont {G.}~\bibnamefont {Brodin}},\ }\bibfield
  {title} {\bibinfo {title} {From extended phase space dynamics to fluid
  theory},\ }\href {https://doi.org/10.1063/1.3496053} {\bibfield  {journal}
  {\bibinfo  {journal} {Phys. Plasmas}\ }\textbf {\bibinfo {volume} {17}},\
  \bibinfo {pages} {102109} (\bibinfo {year} {2010})}\BibitemShut {NoStop}%
\bibitem [{\citenamefont {Wen}\ \emph {et~al.}(2016)\citenamefont {Wen},
  \citenamefont {Bauke},\ and\ \citenamefont {Keitel}}]{Wen:2016:SP}%
  \BibitemOpen
  \bibfield  {author} {\bibinfo {author} {\bibfnamefont {M.}~\bibnamefont
  {Wen}}, \bibinfo {author} {\bibfnamefont {H.}~\bibnamefont {Bauke}},\ and\
  \bibinfo {author} {\bibfnamefont {C.~H.}\ \bibnamefont {Keitel}},\ }\bibfield
   {title} {\bibinfo {title} {Identifying the {Stern-Gerlach} force of
  classical electron dynamics},\ }\href {https://doi.org/10.1038/srep31624}
  {\bibfield  {journal} {\bibinfo  {journal} {Sci. Rep.}\ }\textbf {\bibinfo
  {volume} {6}},\ \bibinfo {pages} {31624} (\bibinfo {year}
  {2016})}\BibitemShut {NoStop}%
\bibitem [{\citenamefont {Wen}\ \emph {et~al.}(2017)\citenamefont {Wen},
  \citenamefont {Keitel},\ and\ \citenamefont
  {Bauke}}]{Wen:2017:PhysRevA.95.042102}%
  \BibitemOpen
  \bibfield  {author} {\bibinfo {author} {\bibfnamefont {M.}~\bibnamefont
  {Wen}}, \bibinfo {author} {\bibfnamefont {C.~H.}\ \bibnamefont {Keitel}},\
  and\ \bibinfo {author} {\bibfnamefont {H.}~\bibnamefont {Bauke}},\ }\bibfield
   {title} {\bibinfo {title} {Spin-one-half particles in strong electromagnetic
  fields: Spin effects and radiation reaction},\ }\href
  {https://doi.org/10.1103/PhysRevA.95.042102} {\bibfield  {journal} {\bibinfo
  {journal} {Phys. Rev. A}\ }\textbf {\bibinfo {volume} {95}},\ \bibinfo
  {pages} {042102} (\bibinfo {year} {2017})}\BibitemShut {NoStop}%
\bibitem [{\citenamefont {Landau}\ and\ \citenamefont
  {Lifshitz}(1980)}]{Landau:Lifshitz}%
  \BibitemOpen
  \bibfield  {author} {\bibinfo {author} {\bibfnamefont {L.~D.}\ \bibnamefont
  {Landau}}\ and\ \bibinfo {author} {\bibfnamefont {E.~M.}\ \bibnamefont
  {Lifshitz}},\ }\href@noop {} {\emph {\bibinfo {title} {The Classical Theory
  of Fields}}}\ (\bibinfo  {publisher} {Butterworth-Heinemann},\ \bibinfo
  {address} {Oxford},\ \bibinfo {year} {1980})\BibitemShut {NoStop}%
\bibitem [{\citenamefont {Semushin}\ and\ \citenamefont
  {Malka}(2001)}]{Semushin:RSI:2001}%
  \BibitemOpen
  \bibfield  {author} {\bibinfo {author} {\bibfnamefont {S.}~\bibnamefont
  {Semushin}}\ and\ \bibinfo {author} {\bibfnamefont {V.}~\bibnamefont
  {Malka}},\ }\bibfield  {title} {\bibinfo {title} {High density gas jet nozzle
  design for laser target production},\ }\href
  {https://doi.org/10.1063/1.1380393} {\bibfield  {journal} {\bibinfo
  {journal} {Rev. Sci. Instrum.}\ }\textbf {\bibinfo {volume} {72}},\ \bibinfo
  {pages} {2961} (\bibinfo {year} {2001})}\BibitemShut {NoStop}%
\bibitem [{\citenamefont {Schmid}\ and\ \citenamefont
  {Veisz}(2012)}]{Schmid:RSI:2012}%
  \BibitemOpen
  \bibfield  {author} {\bibinfo {author} {\bibfnamefont {K.}~\bibnamefont
  {Schmid}}\ and\ \bibinfo {author} {\bibfnamefont {L.}~\bibnamefont {Veisz}},\
  }\bibfield  {title} {\bibinfo {title} {Supersonic gas jets for laser-plasma
  experiments},\ }\href {https://doi.org/10.1063/1.4719915} {\bibfield
  {journal} {\bibinfo  {journal} {Rev. Sci. Instrum.}\ }\textbf {\bibinfo
  {volume} {83}},\ \bibinfo {pages} {053304} (\bibinfo {year}
  {2012})}\BibitemShut {NoStop}%
\bibitem [{\citenamefont {Lorenz}\ \emph {et~al.}(2019)\citenamefont {Lorenz},
  \citenamefont {Grittani}, \citenamefont {Chacon-Golcher}, \citenamefont
  {Lazzarini}, \citenamefont {Limpouch}, \citenamefont {Nawaz}, \citenamefont
  {Nevrkla}, \citenamefont {Vilanova},\ and\ \citenamefont
  {Levato}}]{Lorenz:2019:MRE}%
  \BibitemOpen
  \bibfield  {author} {\bibinfo {author} {\bibfnamefont {S.}~\bibnamefont
  {Lorenz}}, \bibinfo {author} {\bibfnamefont {G.}~\bibnamefont {Grittani}},
  \bibinfo {author} {\bibfnamefont {E.}~\bibnamefont {Chacon-Golcher}},
  \bibinfo {author} {\bibfnamefont {C.~M.}\ \bibnamefont {Lazzarini}}, \bibinfo
  {author} {\bibfnamefont {J.}~\bibnamefont {Limpouch}}, \bibinfo {author}
  {\bibfnamefont {F.}~\bibnamefont {Nawaz}}, \bibinfo {author} {\bibfnamefont
  {M.}~\bibnamefont {Nevrkla}}, \bibinfo {author} {\bibfnamefont
  {L.}~\bibnamefont {Vilanova}},\ and\ \bibinfo {author} {\bibfnamefont
  {T.}~\bibnamefont {Levato}},\ }\bibfield  {title} {\bibinfo {title}
  {Characterization of supersonic and subsonic gas targets for laser wakefield
  electron acceleration experiments},\ }\href
  {https://doi.org/10.1063/1.5081509} {\bibfield  {journal} {\bibinfo
  {journal} {Matter Radiat. at Extremes}\ }\textbf {\bibinfo {volume} {4}},\
  \bibinfo {pages} {015401} (\bibinfo {year} {2019})}\BibitemShut {NoStop}%
\bibitem [{\citenamefont {Popov}\ \emph {et~al.}(2010)\citenamefont {Popov},
  \citenamefont {Rozmus}, \citenamefont {Bychenkov}, \citenamefont {Naseri},
  \citenamefont {Capjack},\ and\ \citenamefont
  {Brantov}}]{Popov:2010:PhysRevLett.105.195002}%
  \BibitemOpen
  \bibfield  {author} {\bibinfo {author} {\bibfnamefont {K.~I.}\ \bibnamefont
  {Popov}}, \bibinfo {author} {\bibfnamefont {W.}~\bibnamefont {Rozmus}},
  \bibinfo {author} {\bibfnamefont {V.~Y.}\ \bibnamefont {Bychenkov}}, \bibinfo
  {author} {\bibfnamefont {N.}~\bibnamefont {Naseri}}, \bibinfo {author}
  {\bibfnamefont {C.~E.}\ \bibnamefont {Capjack}},\ and\ \bibinfo {author}
  {\bibfnamefont {A.~V.}\ \bibnamefont {Brantov}},\ }\bibfield  {title}
  {\bibinfo {title} {Ion response to relativistic electron bunches in the
  blowout regime of laser-plasma accelerators},\ }\href
  {https://doi.org/10.1103/PhysRevLett.105.195002} {\bibfield  {journal}
  {\bibinfo  {journal} {Phys. Rev. Lett.}\ }\textbf {\bibinfo {volume} {105}},\
  \bibinfo {pages} {195002} (\bibinfo {year} {2010})}\BibitemShut {NoStop}%
\bibitem [{\citenamefont {Ji}\ \emph {et~al.}(2014)\citenamefont {Ji},
  \citenamefont {Pukhov}, \citenamefont {Kostyukov}, \citenamefont {Shen},\
  and\ \citenamefont {Akli}}]{Ji:2014:PhysRevLett.112.145003}%
  \BibitemOpen
  \bibfield  {author} {\bibinfo {author} {\bibfnamefont {L.~L.}\ \bibnamefont
  {Ji}}, \bibinfo {author} {\bibfnamefont {A.}~\bibnamefont {Pukhov}}, \bibinfo
  {author} {\bibfnamefont {I.~Y.}\ \bibnamefont {Kostyukov}}, \bibinfo {author}
  {\bibfnamefont {B.~F.}\ \bibnamefont {Shen}},\ and\ \bibinfo {author}
  {\bibfnamefont {K.}~\bibnamefont {Akli}},\ }\bibfield  {title} {\bibinfo
  {title} {Radiation-reaction trapping of electrons in extreme laser fields},\
  }\href {https://doi.org/10.1103/PhysRevLett.112.145003} {\bibfield  {journal}
  {\bibinfo  {journal} {Phys. Rev. Lett.}\ }\textbf {\bibinfo {volume} {112}},\
  \bibinfo {pages} {145003} (\bibinfo {year} {2014})}\BibitemShut {NoStop}%
\bibitem [{\citenamefont {Sylla}\ \emph {et~al.}(2012)\citenamefont {Sylla},
  \citenamefont {Flacco}, \citenamefont {Kahaly}, \citenamefont {Veltcheva},
  \citenamefont {Lifschitz}, \citenamefont {Sanchez-Arriaga}, \citenamefont
  {Lefebvre},\ and\ \citenamefont {Malka}}]{Sylla:2012:PhysRevLett.108.115003}%
  \BibitemOpen
  \bibfield  {author} {\bibinfo {author} {\bibfnamefont {F.}~\bibnamefont
  {Sylla}}, \bibinfo {author} {\bibfnamefont {A.}~\bibnamefont {Flacco}},
  \bibinfo {author} {\bibfnamefont {S.}~\bibnamefont {Kahaly}}, \bibinfo
  {author} {\bibfnamefont {M.}~\bibnamefont {Veltcheva}}, \bibinfo {author}
  {\bibfnamefont {A.}~\bibnamefont {Lifschitz}}, \bibinfo {author}
  {\bibfnamefont {G.}~\bibnamefont {Sanchez-Arriaga}}, \bibinfo {author}
  {\bibfnamefont {E.}~\bibnamefont {Lefebvre}},\ and\ \bibinfo {author}
  {\bibfnamefont {V.}~\bibnamefont {Malka}},\ }\bibfield  {title} {\bibinfo
  {title} {Anticorrelation between ion acceleration and nonlinear coherent
  structures from laser-underdense plasma interaction},\ }\href
  {https://doi.org/10.1103/PhysRevLett.108.115003} {\bibfield  {journal}
  {\bibinfo  {journal} {Phys. Rev. Lett.}\ }\textbf {\bibinfo {volume} {108}},\
  \bibinfo {pages} {115003} (\bibinfo {year} {2012})}\BibitemShut {NoStop}%
\bibitem [{\citenamefont {Nakamura}\ \emph {et~al.}(2010)\citenamefont
  {Nakamura}, \citenamefont {Bulanov}, \citenamefont {Esirkepov},\ and\
  \citenamefont {Kando}}]{Nakamura:2010:PhysRevLett.105.135002}%
  \BibitemOpen
  \bibfield  {author} {\bibinfo {author} {\bibfnamefont {T.}~\bibnamefont
  {Nakamura}}, \bibinfo {author} {\bibfnamefont {S.~V.}\ \bibnamefont
  {Bulanov}}, \bibinfo {author} {\bibfnamefont {T.~Z.}\ \bibnamefont
  {Esirkepov}},\ and\ \bibinfo {author} {\bibfnamefont {M.}~\bibnamefont
  {Kando}},\ }\bibfield  {title} {\bibinfo {title} {High-energy ions from
  near-critical density plasmas via magnetic vortex acceleration},\ }\href
  {https://doi.org/10.1103/PhysRevLett.105.135002} {\bibfield  {journal}
  {\bibinfo  {journal} {Phys. Rev. Lett.}\ }\textbf {\bibinfo {volume} {105}},\
  \bibinfo {pages} {135002} (\bibinfo {year} {2010})}\BibitemShut {NoStop}%
\bibitem [{\citenamefont {Kawata}\ \emph {et~al.}(2014)\citenamefont {Kawata},
  \citenamefont {Nagashima}, \citenamefont {Takano}, \citenamefont {Izumiyama},
  \citenamefont {Kamiyama}, \citenamefont {Barada}, \citenamefont {Kong},
  \citenamefont {Gu}, \citenamefont {Wang}, \citenamefont {Ma},\ and\
  \citenamefont {et~al.}}]{kawata:HPLSE:2014}%
  \BibitemOpen
  \bibfield  {author} {\bibinfo {author} {\bibfnamefont {S.}~\bibnamefont
  {Kawata}}, \bibinfo {author} {\bibfnamefont {T.}~\bibnamefont {Nagashima}},
  \bibinfo {author} {\bibfnamefont {M.}~\bibnamefont {Takano}}, \bibinfo
  {author} {\bibfnamefont {T.}~\bibnamefont {Izumiyama}}, \bibinfo {author}
  {\bibfnamefont {D.}~\bibnamefont {Kamiyama}}, \bibinfo {author}
  {\bibfnamefont {D.}~\bibnamefont {Barada}}, \bibinfo {author} {\bibfnamefont
  {Q.}~\bibnamefont {Kong}}, \bibinfo {author} {\bibfnamefont {Y.~J.}\
  \bibnamefont {Gu}}, \bibinfo {author} {\bibfnamefont {P.~X.}\ \bibnamefont
  {Wang}}, \bibinfo {author} {\bibfnamefont {Y.~Y.}\ \bibnamefont {Ma}},\ and\
  \bibinfo {author} {\bibnamefont {et~al.}},\ }\bibfield  {title} {\bibinfo
  {title} {Controllability of intense-laser ion acceleration},\ }\href
  {https://doi.org/10.1017/hpl.2014.5} {\bibfield  {journal} {\bibinfo
  {journal} {High Power Laser Science and Engineering}\ }\textbf {\bibinfo
  {volume} {2}},\ \bibinfo {pages} {e4} (\bibinfo {year} {2014})}\BibitemShut
  {NoStop}%
\bibitem [{\citenamefont {Park}\ \emph {et~al.}(2019)\citenamefont {Park},
  \citenamefont {Bulanov}, \citenamefont {Bin}, \citenamefont {Ji},
  \citenamefont {Steinke}, \citenamefont {Vay}, \citenamefont {Geddes},
  \citenamefont {Schroeder}, \citenamefont {Leemans}, \citenamefont
  {Schenkel},\ and\ \citenamefont {Esarey}}]{Park:2019:POP}%
  \BibitemOpen
  \bibfield  {author} {\bibinfo {author} {\bibfnamefont {J.}~\bibnamefont
  {Park}}, \bibinfo {author} {\bibfnamefont {S.~S.}\ \bibnamefont {Bulanov}},
  \bibinfo {author} {\bibfnamefont {J.}~\bibnamefont {Bin}}, \bibinfo {author}
  {\bibfnamefont {Q.}~\bibnamefont {Ji}}, \bibinfo {author} {\bibfnamefont
  {S.}~\bibnamefont {Steinke}}, \bibinfo {author} {\bibfnamefont {J.-L.}\
  \bibnamefont {Vay}}, \bibinfo {author} {\bibfnamefont {C.~G.~R.}\
  \bibnamefont {Geddes}}, \bibinfo {author} {\bibfnamefont {C.~B.}\
  \bibnamefont {Schroeder}}, \bibinfo {author} {\bibfnamefont {W.~P.}\
  \bibnamefont {Leemans}}, \bibinfo {author} {\bibfnamefont {T.}~\bibnamefont
  {Schenkel}},\ and\ \bibinfo {author} {\bibfnamefont {E.}~\bibnamefont
  {Esarey}},\ }\bibfield  {title} {\bibinfo {title} {Ion acceleration in laser
  generated megatesla magnetic vortex},\ }\href
  {https://doi.org/10.1063/1.5094045} {\bibfield  {journal} {\bibinfo
  {journal} {Phys. Plasmas}\ }\textbf {\bibinfo {volume} {26}},\ \bibinfo
  {pages} {103108} (\bibinfo {year} {2019})}\BibitemShut {NoStop}%
\bibitem [{\citenamefont {Kulsrud}\ \emph {et~al.}(1986)\citenamefont
  {Kulsrud}, \citenamefont {Valeo},\ and\ \citenamefont
  {Cowley}}]{Kulsrud:1986:Nuclear-Fusion}%
  \BibitemOpen
  \bibfield  {author} {\bibinfo {author} {\bibfnamefont {R.}~\bibnamefont
  {Kulsrud}}, \bibinfo {author} {\bibfnamefont {E.}~\bibnamefont {Valeo}},\
  and\ \bibinfo {author} {\bibfnamefont {S.}~\bibnamefont {Cowley}},\
  }\bibfield  {title} {\bibinfo {title} {Physics of spin-polarized plasmas},\
  }\href {http://stacks.iop.org/0029-5515/26/i=11/a=001} {\bibfield  {journal}
  {\bibinfo  {journal} {Nuclear Fusion}\ }\textbf {\bibinfo {volume} {26}},\
  \bibinfo {pages} {1443} (\bibinfo {year} {1986})}\BibitemShut {NoStop}%
\bibitem [{\citenamefont {Thomas}(1926)}]{Thomas:1926:Motion_of}%
  \BibitemOpen
  \bibfield  {author} {\bibinfo {author} {\bibfnamefont {L.~H.}\ \bibnamefont
  {Thomas}},\ }\bibfield  {title} {\bibinfo {title} {The motion of the spinning
  electron},\ }\href {https://doi.org/10.1038/117514a0} {\bibfield  {journal}
  {\bibinfo  {journal} {Nature}\ }\textbf {\bibinfo {volume} {117}},\ \bibinfo
  {pages} {514} (\bibinfo {year} {1926})}\BibitemShut {NoStop}%
\bibitem [{\citenamefont {Bargmann}\ \emph {et~al.}(1959)\citenamefont
  {Bargmann}, \citenamefont {Telegdi},\ and\ \citenamefont
  {Michel}}]{Bargmann:Telegdi:1959:Precession_of}%
  \BibitemOpen
  \bibfield  {author} {\bibinfo {author} {\bibfnamefont {V.}~\bibnamefont
  {Bargmann}}, \bibinfo {author} {\bibfnamefont {V.~L.}\ \bibnamefont
  {Telegdi}},\ and\ \bibinfo {author} {\bibfnamefont {L.}~\bibnamefont
  {Michel}},\ }\bibfield  {title} {\bibinfo {title} {Precession of the
  polarization of particles moving in a homogeneous electromagnetic field},\
  }\href {https://doi.org/10.1103/PhysRevLett.2.435} {\bibfield  {journal}
  {\bibinfo  {journal} {Phys. Rev. Lett.}\ }\textbf {\bibinfo {volume} {2}},\
  \bibinfo {pages} {435} (\bibinfo {year} {1959})}\BibitemShut {NoStop}%
\bibitem [{\citenamefont {Rakitzis}\ \emph {et~al.}(2003)\citenamefont
  {Rakitzis}, \citenamefont {Samartzis}, \citenamefont {Toomes}, \citenamefont
  {Kitsopoulos}, \citenamefont {Brown}, \citenamefont {Balint-Kurti},
  \citenamefont {Vasyutinskii},\ and\ \citenamefont
  {Beswick}}]{Rakitzis:Science:2003:1936}%
  \BibitemOpen
  \bibfield  {author} {\bibinfo {author} {\bibfnamefont {T.~P.}\ \bibnamefont
  {Rakitzis}}, \bibinfo {author} {\bibfnamefont {P.~C.}\ \bibnamefont
  {Samartzis}}, \bibinfo {author} {\bibfnamefont {R.~L.}\ \bibnamefont
  {Toomes}}, \bibinfo {author} {\bibfnamefont {T.~N.}\ \bibnamefont
  {Kitsopoulos}}, \bibinfo {author} {\bibfnamefont {A.}~\bibnamefont {Brown}},
  \bibinfo {author} {\bibfnamefont {G.~G.}\ \bibnamefont {Balint-Kurti}},
  \bibinfo {author} {\bibfnamefont {O.~S.}\ \bibnamefont {Vasyutinskii}},\ and\
  \bibinfo {author} {\bibfnamefont {J.~A.}\ \bibnamefont {Beswick}},\
  }\bibfield  {title} {\bibinfo {title} {{Spin-Polarized Hydrogen Atoms from
  Molecular Photodissociation}},\ }\href
  {https://doi.org/10.1126/science.1084809} {\bibfield  {journal} {\bibinfo
  {journal} {Science}\ }\textbf {\bibinfo {volume} {300}},\ \bibinfo {pages}
  {1936} (\bibinfo {year} {2003})}\BibitemShut {NoStop}%
\bibitem [{\citenamefont {Sofikitis}\ \emph {et~al.}(2017)\citenamefont
  {Sofikitis}, \citenamefont {Glodic}, \citenamefont {Koumarianou},
  \citenamefont {Jiang}, \citenamefont {Bougas}, \citenamefont {Samartzis},
  \citenamefont {Andreev},\ and\ \citenamefont
  {Rakitzis}}]{Sofikitis:2017PhysRevLett.118.233401}%
  \BibitemOpen
  \bibfield  {author} {\bibinfo {author} {\bibfnamefont {D.}~\bibnamefont
  {Sofikitis}}, \bibinfo {author} {\bibfnamefont {P.}~\bibnamefont {Glodic}},
  \bibinfo {author} {\bibfnamefont {G.}~\bibnamefont {Koumarianou}}, \bibinfo
  {author} {\bibfnamefont {H.}~\bibnamefont {Jiang}}, \bibinfo {author}
  {\bibfnamefont {L.}~\bibnamefont {Bougas}}, \bibinfo {author} {\bibfnamefont
  {P.~C.}\ \bibnamefont {Samartzis}}, \bibinfo {author} {\bibfnamefont
  {A.}~\bibnamefont {Andreev}},\ and\ \bibinfo {author} {\bibfnamefont {T.~P.}\
  \bibnamefont {Rakitzis}},\ }\bibfield  {title} {\bibinfo {title} {Highly
  nuclear-spin-polarized deuterium atoms from the uv photodissociation of
  deuterium iodide},\ }\href {https://doi.org/10.1103/PhysRevLett.118.233401}
  {\bibfield  {journal} {\bibinfo  {journal} {Phys. Rev. Lett.}\ }\textbf
  {\bibinfo {volume} {118}},\ \bibinfo {pages} {233401} (\bibinfo {year}
  {2017})}\BibitemShut {NoStop}%
\end{thebibliography}%

\end{document}